\documentclass[a4paper,fleqn]{cas-dc}

\usepackage[numbers]{natbib}
\usepackage{multirow}
\usepackage{rotating}
\usepackage{physics, amsmath, mathtools}
\usepackage[short,c3]{optidef}
\usepackage{nomencl}
\makenomenclature
\usepackage{multicol}
\usepackage{amsfonts}
\usepackage[mathscr]{eucal}
\usepackage{setspace}
\usepackage{pgf}
\usepackage{pifont}
\usepackage{amssymb}
\newcommand{\xmark}{\ding{55}}
\newcommand{\cmark}{\ding{51}}
\usepackage{latexsym,epsfig}
\usepackage{latexsym,epsfig}
\usepackage{lipsum}
\usepackage{cuted}
\usepackage{caption}
\usepackage{subcaption}
\usepackage{float}
\usepackage{soul}
\usepackage{color}
\usepackage{enumitem}

\usepackage{etoolbox}

\newlength{\nomitemorigsep}
\renewcommand\nomgroup[1]{%
  \itemsep\nomitemorigsep%
  \item[\bfseries
  \ifstrequal{#1}{A}{Acronyms}{%
  \ifstrequal{#1}{D}{Set parameters}{%
  \ifstrequal{#1}{E}{Sets}{%
  \ifstrequal{#1}{F}{Sub-index}{%
  \ifstrequal{#1}{G}{Super-index}{%
  \ifstrequal{#1}{V}{Variables}{%
  \ifstrequal{#1}{W}{Vectors and matrices inputs}{%
  \ifstrequal{#1}{C}{Constants and parameters}{}}}}}}}}%
  ]
 \itemsep\nomitemsep
 }

\let\olditemize\itemize
\def\itemize{\olditemize\itemsep=0pt }

\def\tsc#1{\csdef{#1}{\textsc{\lowercase{#1}}\xspace}}
\tsc{WGM}
\tsc{QE}
\tsc{EP}
\tsc{PMS}
\tsc{BEC}
\tsc{DE}
\RequirePackage[normalem]{ulem} 
\RequirePackage{color}\definecolor{RED}{rgb}{1,0,0}\definecolor{BLUE}{rgb}{0,0,1} 

\begin{document}
\let\WriteBookmarks\relax
\def\floatpagepagefraction{1}
\def\textpagefraction{.001}
\shorttitle{Applied Energy}
\shortauthors{J. Castellanos et~al.}

\title [mode = title]{{An energy management} {system model} with power quality constraints {for unbalanced multi-microgrids interacting in a local energy market} }

\author[1,2]{Johanna Castellanos}[type=editor,
                        orcid=0000-0003-4814-3337]
\cormark[1]
\fnmark[1]
\ead{johanna.castellanos@javeriana.edu.co}
\credit{Conceptualization, Methodology, Software, Writing- Original draft}

\author[1]{Carlos Adrian Correa-Florez}[%
   ]
\address[1]{Pontificia Universidad Javeriana, Electronics Department, Bogota, Colombia}
\ead{carlosa-correaf@javeriana.edu.co}
\credit{Conceptualization, Supervision, Writing - Review \& Editing}

\author[3]{Alejandro Garcés}
\fnmark[2]
\ead{alejandro.garces@utp.edu.co}
\credit{Conceptualization, Writing - Review \& Editing. }

\author[4]{Gabriel Ord\'oñez-Plata}
\fnmark[2]
\ead{gaby@uis.edu.co}
\credit{Resources, Writing - Review \& Editing}

\author [2]{C\'esar A. Uribe}
\fnmark[4]
\ead{cauribe@rice.edu}
\ead[URL]{https://cauribe.rice.edu/}
\credit{Supervision, Writing - Reviewing \& Editing}

\address[2]{Rice University, Electrical and Computer Engineering, Houston, TX, USA}

\address[3]{Universidad Tecnologica de Pereira, Electric Engineering, Pereira, Risaralda, Colombia}

\address[4]{Universidad Industrial de Santander, School of Electric, Electronics and Telecommunications, Bucaramanga, Santander, Colombia}

\author[1]{Diego Patino}
\ead{patino-d@javeriana.edu.co}
\credit{Supervision, Writing - Review \& Editing}

\begin{abstract}
  As {multi-microgrids become readily available, some limited models have been proposed that study operational and power quality constraints with local energy markets independently.} This paper proposes a convex optimization model of an energy management system with operational and power quality constraints and interactions in a  {Local Energy Market (LEM) for unbalanced} microgrids (MGs). {The LEM} consists of a pre-dispatch {step} and an energy transactions step {(ETS)}.  The ETS combines the MGs' objectives while considering two strategies: minimize the cost of buyers or maximize the revenue of sellers. Our proposed model considers harmonic distortion and voltage limit power quality constraints in both steps. Moreover, we model operational constraints such as power flow, power balance, and distributed energy resources behaviors and capacities. We  numerically evaluate the proposed model using three unbalanced MGs with residential, industrial, and commercial load profiles, where each microgrid manages its resources locally. {Furthermore, we create two groups of cases to analyze the interactions in the local energy market. In the first group, the price of the DSO energy and the surplus from MGs to DSO are the same.  The numerical results show that using the increasing revenue strategy promotes MGs to interact more while encouraging them to have high energy prices. When the reducing cost strategy is used, fewer energy interactions occur, and the price of MGs energy is encouraged to be lower.  }
\end{abstract}

\begin{highlights}
\item  {Local Energy Market for energy transactions among unbalanced microgrids and DSO}
\item {Multi-microgrids energy management with operational and power quality constraints }
\item {Unbalanced linear power flow and first-principal-based harmonic power flow modeling}
\end{highlights}


\maketitle 


\mbox{}

\nomenclature[A]{\(\small \text{MG}\)}{\small Microgrid}
\nomenclature[A]{\(\small \text{PV}\)}{\small Photovoltaic}
\nomenclature[A]{\(\small \text{DSO}\)}{\small Distribution System Operator}
\nomenclature[A]{\(\small \text{MMG}\)}{\small Multi-Microgrids}
\nomenclature[A]{\(\small \text{THD}\)}{\small Total Harmonic Distortion}
\nomenclature[A]{\(\small \text{EMS}\)}{\small Energy Management System}
\nomenclature[A]{\(\small \text{BESS}\)}{\small Battery Energy Storage System}
\nomenclature[A]{\(\small \text{SoC}\)}{\small State of Charge}

\nomenclature[D]{\(g\)}{\small Number of MGs considered}
\nomenclature[D]{\(n_i\)}{\small Number of nodes of MG $i$}

\nomenclature[E]{\(\mathbb{R}\)}{\small Real numbers}
\nomenclature[E]{\(\mathbb{C}\)}{\small Complex numbers}
\nomenclature[E]{\(\mathbb{N}\)}{\small Natural numbers}
\nomenclature[E]{\small \(\mathcal{M}=\{1,2,\dots,g\} \)}{\small Microgrids}
\nomenclature[E]{\small \(\mathcal{F}=\{a,b,c\}\)}{\small Phases}
\nomenclature[E]{\small \(\mathcal{T}=\{1,\dots,24\}\)}{\small Time}
\nomenclature[E]{\small \(\mathcal{H}=\{3,5,7,11,13 \}\)}{\small Harmonic frequencies}
\nomenclature[E]{\small \(\mathcal{A}=\{\mathcal{M} \cup 0 \}\)}{\small DSO and Microgrids}
\nomenclature[E]{\small \(\mathcal{N}_i=\{1,2,\dots,n_i \}\)}{\small Nodes of the MG $i$}

\nomenclature[F]{\small \( i \in \mathcal{M} \)}{\small Microgrid}
\nomenclature[F]{\small \( j \in \mathcal{A} \)}{\small Microgrid or DSO}
\nomenclature[F]{\small \( k,m \in \mathcal{N}_i\)}{\small Nodes}
\nomenclature[F]{\small \( \phi \in \mathcal{F} \)}{\small Phase}
\nomenclature[F]{\small \( 3\phi  \)}{\small Three-phase. It is not 3 times $\phi$}
\nomenclature[F]{\(h \in \mathcal{H}\)}{\small Harmonic frequency}

\nomenclature[G]{\small \( G  \)}{\small Generation}
\nomenclature[G]{\small \( D  \)}{\small Demand}
\nomenclature[G]{\small \( L  \)}{\small Loads}
\nomenclature[G]{\small \( pv \)}{\small PV systems}
\nomenclature[G]{\small \( bd  \)}{\small Battery discharge}
\nomenclature[G]{\small \( bc  \)}{\small Battery charge}
\nomenclature[G]{\small \( b  \)}{\small Battery}

\nomenclature[W]{\small \( [V_i(t)]_{k} = v_{i,k}(t)\)}{\small Voltage in node $k \in \mathcal{N}$ of MG $i$ at time~$t$}
\nomenclature[W]{\small \( [V_{i,\phi}(t)]_{k} = v_{i,\phi,k}(t)\)}{\small Voltage in node $k$ per phase $\phi$ of MG $i$ at time~$t$}
\nomenclature[W]{\small \( V_{i,3\phi}(t) = [V_{i,a}(t) \hspace{4pt} V_{i,b}(t) \hspace{4pt} V_{i,c}(t)]^{\operatorname{T}} \)}{\small Nodal voltages per phase $\phi$ of MG $i$ at time $t$} 
\nomenclature[W]{\small \( [V_{l_i}]_{k0} = v_{i,k0}\)}{\small Linearization voltage in node $k$ of MG $i$}
\nomenclature[W]{\small \( [I_i(t)]_{k} = \text{i}_{i,k}(t)\)}{\small Current in node $k$ of MG $i$ at time $t$}
\nomenclature[W]{\small \( [I_{i,\phi}(t)]_{k} = \text{i}_{i,\phi,k}(t)\)}{\small Current in node $k$ per phase $\phi$ of MG $i$ at time~$t$}
\nomenclature[W]{\small \( [I_{i,3\phi}(t)]_{\phi} = I_{i,\phi}(t)\)}{\small Nodal currents per phase $\phi$ of MG $i$ at time~$t$}
\nomenclature[W]{\small \( [S_i(t)]_{k} = s_{i,k}(t) \)}{\small Power in node $k$ of MG $i$ at time~$t$}
\nomenclature[W]{\small \( [S_{i,\phi}(t)]_{k} = s_{i,\phi,k}(t)\)}{\small Power in node $k$ per phase $\phi$ of MG $i$ at time~$t$}
\nomenclature[W]{\small \( [S_{i,3\phi}(t)]_{\phi} = S_{i,\phi}(t)\)}{\small Nodal powers vector per phase $\phi$ of MG $i$ at time $t$}
\nomenclature[W]{\small \( [Y_i]_{km} = y_{i,km} \)}{\small Admittance from node $k$ to $m$ of MG $i$}
\nomenclature[W]{\small \( [\overline{P}^{pv}_{i}(t)]_{k} =  \overline{p}^{pv}_{i,k}(t) \)}{\small PV system capacity in node $k$ of MG~$i$ at time $t$}
\nomenclature[W]{\small \( [\overline{P}^{pv}_{i,\phi}(t)]_{k} = \overline{p}^{pv}_{i,\phi,k}(t)\)}{\small PV system capacity in node $k$ per phase $\phi$ of MG~$i$ at time $t$}
\nomenclature[W]{\small \( [\overline{P}^{pv}_{i,3\phi}(t)]_{\phi} = \overline{P}^{pv}_{i,\phi}(t)\)}{\small PV system capacity vector per phase $\phi$ of MG $i$ at time $t$}
\nomenclature[W]{\small \( [P^{pv}_{i}(t)]_{k}=p^{pv}_{i,k}(t) \)}{\small PV system active power in node $k$ of MG $i$ at time $t$}
\nomenclature[W]{\small \( [P^{pv}_{i,\phi}(t)]_{k} = p^{pv}_{i,\phi,k}(t)\)}{\small PV system active power in node $k$ per phase $\phi$ of MG $i$ at time $t$}
\nomenclature[W]{\small \( [P^{pv}_{i,3\phi}(t)]_{\phi} = P^{pv}_{i,\phi}(t)\)}{\small PV system active power vector per phase $\phi$ of MG $i$ at time $t$}

\nomenclature[C]{\(\small \zeta_{j}(t)\)}{\small Energy price of agent $j$ at time $t$}
\nomenclature[C]{\(\small \Hat{\zeta}_{0}(t)\)}{\small Surplus energy price from MGs to DSO at time $t$}
\nomenclature[C]{\(Y_{i,3\phi}\)}{\small Three-phase admittance matrix of the MG $i$ network}

\nomenclature[C]{\(S_{i}^{L}(t)\)}{\small Loads power vector of MG $i$ at time $t$}
\nomenclature[C]{\(S_{i,3\phi}^{L}(t)\)}{\small 3-phase loads power vector of MG $i$ at time $t$}

\nomenclature[V]{\(\small \rho_i(t)\)}{\small Purchase cost in the EMS of MG $i$ at time $t$}
\nomenclature[V]{\(\small \Upsilon_i(t)\)}{\small Energy market revenue of MG $i$ at time $t$}

\nomenclature[V]{\(\small \mathscr{P}_{ij}(t)\)}{\small Energy sold in EMS from MG $i$ to agent $j$ at time $t$}

\nomenclature[V]{\small \(V_{i,h}(t)\)}{\small Nodal voltages vector of MG $i$ at harmonic $h$  and time $t$}
\nomenclature[V]{\small \(V_{i,h,\phi}(t)\)}{\small Nodal voltages vector of MG $i$ at harmonic $h$, phase $\phi$ and time $t$}
\nomenclature[V]{\small \(I_{i,3\phi}(t)\)}{\small Three-phase nodal currents vector of MG $i$ at time $t$}

\nomenclature[V]{\small \(S_{i,h}(t)\)}{\small Nodal powers vector of MG $i$ at harmonic $h$  and time $t$}
\nomenclature[V]{\small \(S_{i,h,\phi}(t)\)}{\small Nodal powers vector of MG $i$ at harmonic $h$, phase $\phi$ and time $t$}

\nomenclature[V]{\small \(S_{i}^G(t)\)}{\small Generation power vector of MG $i$ at time $t$}
\nomenclature[V]{\small \(S_{i,3\phi}^G(t)\)}{\small 3-phase generation power vector of MG $i$ at time $t$}

\printnomenclature


\section{Introduction}

 {Microgrids (MGs) are often smart-grid-inspired local infrastructures that run independently of the {main} power system and group numerous individuals within a local structure \mbox{\cite{Sumper_19}}\hspace{0pt}. As a result, the idea of "Multi-Microgrids" (MMG) emerges, representing the connections between MGs in terms of power and energy.  
 }

 {Multi-Microgrids are structures} conformed by low-voltage MGs and distributed generation units connected to adjacent medium-voltage feeders.  In this sense, many low-voltage networks, with micro-sources and loads, are now active elements and must be coordinated. However, this increases the complexity of designing energy management protocols  \cite{Gonzales-Zurita_20, Du_18}.

  Microgrids' energy components are scheduled by the energy management system (EMS), which also decides how they will interact with the main grid. To achieve an affordable, sustainable, and {reliable} operation of the MG, {the EMS} must consider supply and demand management while fulfilling system constraints.  Moreover, the EMS is responsible for coordinating the power flows from all the involved agents \mbox{\cite{Zia_18, Khavari_20}}.

  Microgrid operator aims to maximize local resources to use the least amount of energy from suppliers. Thus, power flow modeling is necessary. Power flow is a non-convex phenomenon. It is usually {considered in optimization problems }via convex second-order cones, semi-definite programming~\cite{Madureira_10, Haghifam_20, Wang_Yu_20, Arkhangelski_21, Farahani_17, GaoH_18, Yan_21, Wang_Yifei_20, Zhang_17, Karim_17, Baghaee_18, Yan_21}, or surrogate linear constraints \cite{Du_18, Mo_21, Pinzon_19, Marini_20, Zheng_18}.  Non-linear or software-based power flow models have also been explored in the literature~\cite{Carpinelli_17, Bazmohammadi_20, Amir_17, Thomas_20, Nazari_21}. {Furthermore, MG energy management systems can include balanced or unbalanced power flow. In the case of unbalanced PF, the three-phased lines are modeled with different power loads per phase \cite{Ramirez-Garces_19}.}

   Electricity markets are marketplaces for selling and purchasing energy. In this context, a local power market is defined as a geographical area inside a local energy community that offers energy or flexibility as part of a public grid \cite{Ampatzis_14}.  As a result, two local markets can be considered: local energy markets (LEMs) and local flexibility markets (LFMs), which can be centralized or peer-to-peer (P2P) \cite{Sumper_19}.  Local power markets play a significant role in MMG energy transactions providing economic incentives for expanding generation capacity near load centers \cite{Staudt_17}. This type of market is controlled by a market operator, whose role may vary based on the specific electrical market, such as the day-ahead market, intraday, forwards, etc, where the goal is to solve dispatch locally.  \mbox{\cite{Sumper_19}.} LEM seeks to reduce power consumption from the main grid through local energy community trading of energy produced from local resources \mbox{\cite{Mengelkamp_17, EuropCom_16, Ilieva_16}}.

  {As the number of MGs available grows, operational and Power Quality (PQ) requirements emerge~\cite{Alkahtani_20,doe_17}. The system requires interconnections that provide economic benefits to MGs and DSOs and keep the distribution grid and the MGs operational. Among the most common PQ issues are signal distortion, frequency, voltage unbalance, transients, voltage sags/swells, bidirectional power flows, and short-term interruptions~\cite{Alkahtani_20,Luo_16}
  }.   
  
  {We focus on signal distortion and over-voltage events, two of the most prevalent PQ problems in MGs operating in grid-connected mode.} Signal distortion can be induced by non-linear loads, {inverter switching mechanisms,} power electronics devices, {inverter position} in the grid, number of inverters in the network, and  {variations in solar irradiation~\cite{Anurangi_17, Karimi_16,Anzalchi_19}. Signal distortion bounds can be guaranteed by including Total Harmonic Distortion (THD) constraints~\cite{Karim_17, Marini_20}. However, to acquire the THD, harmonic voltages must be obtained via} Harmonic Power Flow (HPF). {There are two ways to obtain HPF in {EMS} formulations: software-based~\cite{Thomas_20,Nazari_21}, and first-principles based } models~\cite{Marini_20, Karim_17, Baghaee_18}. The downside of software-based models is that they use heuristic techniques. Thus, achieving optimal energy management can not be guaranteed~\cite{Karimi_16,Anzalchi_19}.

  We present  an incomplete list of existing {research } with energy trading formulations and PQ constraints {for {an MG} in Table~\ref{tab:ref-pq}.  We focus on three types of PQ constraints: voltage limits, harmonic distortion, and voltage {unbalance}}.  Moreover, we identify the presence of PF and HPF formulations.

    \begin{table}[htb] \footnotesize
    \caption{Research with Energy Tradings (ET) including PQ problems, PF and HPF formulations without MGs interactions. \\
    {\footnotesize {EMS=Energy Management Systems},  {EMS-PM} ={Enegy Management System with Price Mechanism} , {LFM = Local Flexibility Markets} {PASM} = {Power Ancillary Services } Market, VL=Voltage Limits, {VM=Voltage Management, } VC=Voltage Control, HD=Harmonic Distortion, U=Unbalanced, CPF=Convex Power Flow, OPF=Optimal Power Flow, NLPF=Non-linear Power Flow } }\label{t:ref-pq}
    \begin{tabular}{ m{2.9cm}CCCC } 
    \toprule
     Reference & ET & PQ  & PF & HPF \\
    \midrule
         {\textit{{Pinzon,2019}}\mbox{\cite{Pinzon_19}}} & {EMS }& {VL }& {CPF }& \xmark \\ 
        \textit{{Paudel,2021}}\mbox{\cite{Paudel_21}  } & {EMS-PM }& {VL-VM }& {IPF }& \xmark \\ 
         \textit{Arkhangelski,2021} \cite{Arkhangelski_21}, {\hspace{1em} } \textit{Carpinelli,2017} \cite{Carpinelli_17}  & {EMS} & VL,U & NLPF & \xmark \\ 
        \textit{ {Marini} , {2020} } {\mbox{\cite{Marini_20} }} & {EMS} & VL,HD  & CPF &  \cmark  \\ 
        {\textit{ Thomas ,2020}  \mbox{\cite{Thomas_20} }} & {EMS} & VL,HD & IPF  & \cmark  \\
        \textit{{Nazari,2021}} {\mbox{\cite{Nazari_21} }}& {EMS-PM }& {VL,HD }& {IPF }& \cmark \\
        \textit{{Hong,2015}} {\mbox{\cite{Hong_15} }}& {EMS }& {VC,HD,U }& \xmark & \xmark \\ 
         \textit{Karim,2017} \cite{Karim_17} & {EMS} & \hspace{0.5em} VL,HD,U \hspace{0.5em} & CPF & \cmark \\ 
        \textit{Baghaee,2018} \cite{Baghaee_18}  & {EMS} & VL,HD,U & PF & \cmark \\
       \textit{Farahani,2017} \cite{Farahani_17}  & {LFM-PASM } & HD & OPF & \xmark  \\ 
    \bottomrule
    \end{tabular}
    \label{tab:ref-pq}
    \end{table}

 In \mbox{\cite{Baghaee_18}}, energy management with HPF with non-linear loads and voltage unbalance is proposed for a microgrid with balanced and unbalanced operations.  In \mbox{\cite{Marini_20}}, the authors present a model for harmonic compensation through energy storage systems, including a linear power flow and HPF. 

  Except for \mbox{\cite{Farahani_17}}, whose work focuses on local auxiliary services of harmonic distortion and reactive power, most publications often include voltage limit constraints. \mbox{\cite{Thomas_20, Marini_20, Nazari_21} } include the combination of harmonic distortion and voltage limits constraints. In addition, the combination of signal distortion, voltage unbalance, and voltage limits constraints are modeled in \mbox{\cite{Hong_15,Karim_17,Baghaee_18}}.

   Previous works incorporate PQ, including HPF, with principle-based models. However, the goal of these energy trades is the dispatch of a single MG. Therefore, there are no energy exchanges with other agents, e.g., other MGs.

  {Table \ref{tab:ref-MGs} summarizes available energy trading formulations, including interactions between MGs and DSO. The four categories we use to classify energy trading are {Energy Management System (EMS)}, Aggregator (A), Local Energy Market (LEM), and } Transactive Energy Schemes (TES). {Transactive energy (TE) enables peers to trade energy}. {A transactive energy framework is made up of several interconnected agents, including a market for energy, service providers, generation firms, transmission and distribution networks, prosumers, along others~\cite{Siano_19, Ghamkhari_19}.  LEMs, however, can potentially empower {small} players and serve as the first step toward fully transactive energy systems \mbox{\cite{Lezama_19}}.
  } 

    \begin{table*}[htb] \footnotesize
    \caption{Research associated with MGs interactions contrasting with a type of energy trading (ET), PQ problems, power flow (PF), and harmonic power flow (HPF) formulations.\\
    {\footnotesize MMG= Multi-Microgrids, DSO=Distribution System Operator, {EMS=Energy Management System}, A=Agregator,   LEM=Local Energy Market, {TES=Transactive Energy Scheme, } VL=Voltage Limits, IL=Current limits, VC=Voltage Control,  HD=Harmonic Distortion, U={Voltage unbalanced} , CPF=Convex Power Flow, OPF=Optimal Power Flow, IPF=Iterative Power Flow, NLPF=Non-linear Power Flow} }\label{t:comp}
     \begin{tabular}{ m{7.4cm} C m{1.6cm} C m{0.8cm} C } 
     \toprule
     Reference & Interaction & ET  & PQ  & PF & HPF \\
    \midrule
   \textit{Wang,2018} \cite{Wang_18} &{ MMG-DSO} & \xmark & VC & \xmark & \xmark \\
       \textit{Nikmehr,2015}\cite{Nikmehr_15},  \textit{Amir,2017}\cite{Amir_17}, \textit{Bazmohammadi,2019}\cite{Bazmohammadi_19},  \textit{Movahednia,2020} \cite{Movahednia_20},   \textit{Qiu,2020} \cite{Qiu_20_Robustly},   \textit{Sang,2020} \cite{Sang_20}, \textit{Yang,2020} \cite{Yang_Xiaodong_20},  \textit{Zhu,2020} \cite{Zhu_20}  & MMG & {EMS} & \xmark & \xmark & \xmark  \\  
       \textit{Qiu,2020} \cite{Qiu_20_Decentralized}, \textit{Karimi,2021} \cite{Karimi_21_Stochastic} & MMG-DSO & {EMS} & \xmark & \xmark & \xmark  \\  
        \textit{Zhang,2017} \cite{Zhang_17} & MMG & {EMS} & \xmark & OPF & \xmark \\   
        \textit{GaoH,2018} \cite{GaoH_18} & MMG-DSO & {EMS} & \xmark & CPF & \xmark   \\  
        \textit{Bayat,2020} \cite{Bayat_20}, \textit{Esmaeili,2020} \cite{Esmaeili_20}   & \hspace{0.5em} MMG-DSO \hspace{0.5em} & {EMS}  & VL & \xmark & \xmark \\  
        \textit{Wang,2020} \cite{Wang_Yu_20} & MMG & {EMS} & VC & OPF & \xmark \\   
        \textit{Bazmohammadi,2020} \cite{Bazmohammadi_20} & MMG & {EMS} & VL, IL & IPF & \xmark \\   
        \textit{Mo,2021} \cite{Mo_21} & MMG & {EMS} & VL & CPF & \xmark \\  
        \textit{Du,2018}\cite{Du_18} & MMG-DSO & {EMS} & VL & CPF & \xmark \\   
        \textit{Khavari,2019}\cite{Khavari_19}, \textit{Khavari,2020} \cite{Khavari_20} & MMG & A & \xmark & \xmark & \xmark \\   
         \textit{Haghifam,2020} \cite{Haghifam_20} & MMG-DSO &  A & VL & OPF & \xmark \\  
        \textit{Madureira,2010}\cite{Madureira_10} & MMG-DSO & LEM & VL,VC & NLPF & \xmark \\   
        \textit{Sheikhahmadi,2020} \cite{Sheikhahmadi_20} & MMG-DSO & LEM & \xmark & \xmark & \xmark \\   
        \textit{Sheikhahmadi,2022} \cite{Sheikhahmadi_22} & MMG & LEM & \xmark & \xmark & \xmark \\   
        \textit{Lezema,2019} \cite{Lezama_19}  & MMG & LEM-TES & \xmark & \xmark & \xmark  \\  
        \textit{Karimi,2021} \cite{Karimi_21_Dynamic} & MMG-DSO & TES & \xmark & \xmark & \xmark  \\  
        \textit{Wu,2020} \cite{Wu_20} & MMG-DSO & TES & \xmark & \xmark & \xmark \\  
        \textit{Wang,2020}\cite{Wang_Yifei_20},\textit{Yan,2021}\cite{Yan_21} & MMG-DSO & TES & VL & CPF & \xmark \\ 
        \rowcolor{lightgray} \textcolor{BlueViolet}{\textbf{This paper}} & \textcolor{BlueViolet}{\textbf{MMG-DSO}} & \textcolor{BlueViolet}{ \textbf{\textbf{{EMS-LEM}} }} & \textcolor{BlueViolet}{\textbf{VL,HD,U}} & \textcolor{BlueViolet}{\textbf{CPF}} & \textcolor{BlueViolet}{\cmark} \\
    \bottomrule
    \end{tabular}
    \label{tab:ref-MGs}
    \end{table*}

  A bi-level Transactive Energy Scheme (TES) approach for MMG is proposed in \mbox{\cite{Karimi_21_Dynamic, Wang_Yifei_20} }where the upper level represents the DSO while the lower level, the MGs. The work in~\cite{Karimi_21_Dynamic} establishes a distributed end-to-end transactive energy trading and pricing model using Lagrangian relaxation and decomposition methods. MGs decide on their own and submit to the DSO their equivalent loads. DSO reconfigures the distribution network to find the transactive energy path among MMG. The lower level will consider TE trading adjustments if distribution network constraints cannot be guaranteed.

  The PQ constraints are not included in the TES formulation in~\cite{Karimi_21_Dynamic}, but they are in the works in~\cite{Wang_Yifei_20, Yan_21}, particularly the voltage limit constraints.  The authors of~\cite{Yan_21} use a Stackelberg game to enable Peer-to-peer (P2P) transactive energy trading while protecting the agents' privacy and managing heterogeneous DER. Furthermore, power network constraints and DSO operations, such as network reconfiguration, are considered.

  Local Energy Markets (LEMs) are proposed in~\cite{Madureira_10, Sheikhahmadi_20, Sheikhahmadi_22, Farahani_17}, where the MGs energy markets offer ancillary services such as voltage or frequency regulation \mbox{\cite{Madureira_10}}, and reactive power compensation~\cite{Madureira_10, Farahani_17}.

  PQ constraints are considered in summary, identifying the presence of PF and HPF in the formulation. Upon closer inspection, many MMG energy trading formulations, such as the models proposed in \mbox{\cite{Du_18, Wang_Yifei_20,Bazmohammadi_20, Mo_21, Haghifam_20, Yan_21}}, include PF and PQ constraints. However, the most common type of PQ constraint is one related to voltage limits or voltage control. No existing LEM or TES models include operational and PQ constraints with voltage {unbalance}, signal distortion, and HPF models.

  {We propose a Multi-Microgrid local market framework involving MGs and DSO composed of two steps and based on a day-ahead centralized control mode with distributed decision-making. The first step is based on an {EMS} for each microgrid; the role of the seller or buyer is determined, as well as the selling or buying energy capacity. The second step is centralized and defines the microgrid interactions. Furthermore, the proposed optimization model includes operational and PQ constraints (unbalanced microgrids, voltage limits, and signal distortion simultaneously).  { The term \textit{ \textbf{unbalanced microgrids}} refers to three-phase microgrids with power differences between phases due to the nature of the LV networks \cite{Sepehry_19}.} Additionally, we use linear PF and a first-principles-based HPF.
 }

  Next, we state the main contributions of this work.

\begin{itemize}
    \item We propose a mathematical model to depict the energy interactions between unbalanced microgrids and DSO {considering} operational and power quality constraints.  The operational constraints consist of {unbalanced} power flow, power balance, DER dynamics, and energy interactions. Moreover, we approximate the non-convex power flow constraints by linearization around operating points.
    \item We extend existing approaches by incorporating the following power quality constraints into the model: harmonic power flow for {unbalanced} MGs, voltage limits, and total harmonic distortion constraints. The THD and voltage limit constraints are non-convex. We use surrogate convex expressions for voltage limits and THD.
    \item We design a local energy market for MGs interactions with two steps: 1. Pre-dispatch and 2. Energy transactions.
\end{itemize}

We validate the performance of the model with numerical analysis. We simulate a test system based on a modified European CIGRE LV benchmark. We add PV systems and batteries where the residential, industrial, and commercial sub-networks are considered {unbalanced} MGs. 

The remainder of the paper is organized as follows. In Section \ref{S2-problem}, we define the problem.  Section 3  {establishes } the operational and PQ constraints, respectively.  Section {4 } presents the proposed market framework and optimization model of the individual and MMG energy management system. Section \ref{S6-sim&res} presents the test system, define simulation cases, and shows the numerical results. Finally, Section \ref{S7-Conc} provides conclusions and future works.

\section{Problem statement} \label{S2-problem}

   This section introduces the energy management system model and proposes local and collective objective functions based on minimizing energy costs. We will start with some useful definitions.  

   Let  $\mathcal{M} = \{i \in \mathbb{N} \mid i=1,2,\dots,g \}$ be the set of unbalanced MGs. A microgrid can be modeled as a graph $\mathcal{G}_i=\{ \mathcal{N}_i, \mathcal{E}_i \}$ with $\mathcal{N}_i$ being the set of all nodes of the microgrid $i \in \mathcal{M}$ and $\mathcal{E}_i = \{(k,m) \in \mathcal{N}_i \times \mathcal{N}_i \} $ being the set of edges or power lines.  Each microgrid $i \in \mathcal{M}$ can interact with other agents $j \in \mathcal{A} = \{ \mathcal{M} \cup 0 \}$, where $\mathcal{A}$ is the set of MGs, and the index zero represents the DSO. We will use the term agent when referring to MGs and DSO.  Moreover, let $\phi \in \mathcal{F}$ be the set of three phases for all MGs in $\mathcal{M}$ where $\mathcal{F} =\{ a, b, c\}$.  Additionally, to include the intertemporal operation, we define the time set  $\mathcal{T}=\{t \in \mathbb{N} \mid 1 \leq t \leq 24\}$ which corresponds to the daily operation over 24 hours.

\begin{figure}[t]
     \centering
     \includegraphics[width=\linewidth]{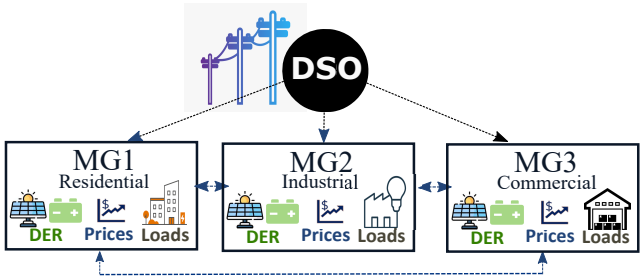}
     \caption{Energy interactions scheme for four agents. MGs and DSO are electrically connected at the Point of Common Coupling (PCC). The parameters of each MG are their load demand, DER, and energy prices which change per hour.}
     \label{fig:interactions}
 \end{figure}

   Figure \ref{fig:interactions} shows a schematic visualization of energy interactions for four agents: three MGs through their own MG operator and the DSO.  Later in Section \ref{S6-sim&res}, we will use this configuration for numerical analysis. 

  Microgrids can share active energy among themselves and the DSO in a local energy market.  The agents electrically connect to the Point of Common Coupling (PCC). Moreover,  we seek to develop an optimization model to minimize the MGs' energy cost interactions in the local energy market. Additionally, the MGs revenue for selling energy needs to be included. Moreover, the model considers two constraints: 1) operational and 2) power quality constraints. 

  The profit of each MG, i.e., $\pi_i(t)$, is given by the difference between revenue and cost, and it can be written as
    \begin{align} \label{eq:obj}
      \pi_i(t)  = \Upsilon_i(t) - \rho_i(t) ,
    \end{align}
  where $\Upsilon_i(t) \in \mathbb{R}$ is the revenue of microgrid $i \in \mathcal{M}$, from energy sales, and $\rho_i(t) \in \mathbb{R}$ represents the total cost of energy purchased from the energy market, both at time $t \in \mathcal{T}$.

  The energy cost of MG $i$, i.e., $\rho_i(t)$, is calculated as the summation of the product between the energy price of the agents $i \neq j$, i.e., $\zeta_{j}(t)$, and the active power, i.e., $\mathscr{P}_{ji}(t)$, that agent $j \in \mathcal{A}$ is selling to MG $i \in \mathcal{M}$ at time $t \in \mathcal{T}$. Thus,  $\rho_i(t)$ is the cost of purchasing energy for every MG and can be written as
      \begin{align}
        \rho_i(t) =  \sum_{\substack{j \in \mathcal{A}\\ i \neq j}} {\zeta_{j}(t) \mathscr{P}_{ji}(t)}. \label{eq:rho}
    \end{align}

  The total revenue per MG is composed of the revenue coming from energy sales $\Upsilon_i(t) \in \mathbb{R}$ as shown next:
    \begin{align}
        \Upsilon&_i(t) = \Hat{\zeta}_{0} \mathscr{P}_{i0}+ \sum_{\substack{j \in \mathcal{M}\\ j \neq i}}{\zeta_{i}(t)\mathscr{P}_{ij}(t) }, \label{eq:rev-e} 
    \end{align}
  where $\Upsilon_i(t)$ has two components. The first one is the revenue from energy surplus from MGs to DSO $\mathscr{P}_{i0}$ at price $\Hat{\zeta}_{0}$.  Notice that agent $j=0$, the DSO, is the only agent with a different price for purchasing and selling energy. When MGs deliver energy to the DSO, it is not considered a sale but a surplus delivery. The second one is the revenue obtained when energy is sold to other MGs, in which $\mathscr{P}_{ij}(t)$ is the energy sold from MG $i \in \mathcal{M}$ to $j \in \mathcal{M}$ at a price $\zeta_{i}(t)$ and time $t \in \mathcal{T}$.  

   The goal of each MG $i$ is to maximize its profit, and it can be written as a minimization problem of the cost minus the revenue, i.e.,
  \begin{align} \label{eq:obj1}
        {\min_{\{\mathscr{P}_{ji}(t)\}}} \; \sum_{t \in \mathcal{T} } \Bigg( \overbrace{\sum_{\substack{j \in \mathcal{A} \\ i \neq j}}{\zeta_{j}(t) \mathscr{P}_{ji}(t)}}^{\text{Purchase cost}} \; \overbrace{- \; \; \widehat{\zeta}_{0}(t)\mathscr{P}_{i0}(t)}^\text{Surplus revenue} \nonumber \\ 
       \overbrace{ - \sum_{\substack{j \in \mathcal{M} \\  i \neq j }} {\zeta_{i}(t)\mathscr{P}_{ij}(t)}}^{\text{Sales revenue}} \Bigg).
    \end{align}

    Next, we define a graph $\mathcal{G = \{ \mathcal{A}, \mathcal{E} \}}$, which will allow us to model the MGs interactions. The set of nodes of the graph,  $\mathcal{A}$, corresponds to the set of the agents, and $\mathcal{E} = \{(j,i) \in \mathcal{A} \times \mathcal{A} \} $ to the set of edges.  Note that $\mathcal{G}_i$ represents the network of MG $i$ while $\mathcal{G}$ is the network of MGs. Therefore, we formulate the MMG energy management system problem as
    \begin{align}
        {\min_{\mathscr{P}}} \quad  \sum_{t \in \mathcal{T} } \Bigg( \zeta_{0}(t) \sum_{i \in \mathcal{M}}{\mathscr{P}_{0i}(t)} \; - \; \widehat{\zeta}_{0} \sum_{i \in \mathcal{M}} \mathscr{P}_{i0}
        \nonumber\\ + \sum_{e \in \mathcal{E}} {\zeta_{e}(t)\mathscr{P}_{e}(t)}   \Bigg) \\
        \mathscr{P} = \{\mathscr{P}_{0i}(t), \mathscr{P}_{i0}(t), \mathscr{P}_{e}(t)\}, \nonumber \\
        \forall i \in \mathcal{M}, \; \forall e \in \mathcal{E}, \; \forall t \in \mathcal{T} \nonumber
    \end{align}
    \hspace{0.9cm} subject to

    \begin{enumerate}
        \item Operational constraints:
        \begin{itemize}
            \item \textit{Power flow:}  The variables in the power flow model are nodal net injected power and nodal voltages, both in the complex domain. The power flow constraints are generally non-convex. Thus, we linearized the model around the operational points using Wirtinger derivatives. 
        \item  \textit{Power balance:}  The net injected power is the difference between generation and demand power. Besides, the generation power is the sum of PV system generation, batteries discharge power, and the external generation that satisfies the demand of each microgrid $i$ $\in \mathcal{M}$. Furthermore, demand power is composed of load power and battery charging power.
        \item \textit{Distributed energy resources:}  We consider PV systems and batteries. The maximum generation capacity constrains the power of the PV systems.  This maximum capacity is modeled as a function that depends on temperature effects, inverter efficiency, and time.  Moreover, we model the dynamics of the battery's energy, called State of Charge (SoC), with three variables: SoC, charge, and discharge power.  These variables are restricted by their maximum capacities.  Moreover, the maximum DER capacities are integrated into the model through box constraints.
        \item \textit{Interactions:} The interaction constraints allow the exchange of energy among agents in the market through the MMG energy management system.
        \end{itemize}

        \item PQ constraints:
        \begin{itemize}
            \item \textit{Harmonic power flow:}  We model a convex HPF using the current injection technique \cite{Grady_83}. We include the HPF as an equality constraint where the variables are the harmonic voltages of each harmonic frequency.
            \item \textit{Harmonic distortion:} We use the THD as a metric of the harmonic distortion. We include the THD as a convex inequality constraint. 
            \item \textit{Voltage limits:} We use a convex inequality constraint to the lower and upper bounds of the magnitude of the nodal voltages. 
        \end{itemize}
    \end{enumerate}

 We seek to model the interacting microgrids' active power.  Moreover, we explicitly characterize the operational and PQ constraints to find optimal solutions via computational optimization methods.  

\section{Operational {and power quality} constraints} \label{s:OIC}
This section describes the operational constraints for local and MMG energy management. Next, we present a convex power flow formulation, power balance, DER, and interaction constraints.

\subsection{Power flow}
  \subsubsection{From non-convex to linear power flow}

   The complex voltages $v_k \in \mathbb{C}$ and currents $\mathrm{i}_k \in \mathbb{C}$ are defined per each node $k \in \mathcal{N}_i$ of a microgrid $i \in \mathcal{M}$. For simplicity of notation, we will ignore the microgrid index, phase notation, and time.  In this way, the nodal net complex power $s_k \in \mathbb{C}$ can be defined as
    \begin{align} 
          s^*_k = v^*_k \mathrm{i}_k, 
    \end{align}
 where $(\cdot)^*$ denotes the complex conjugate.  In addition, the complex nodal current is defined as
    \begin{align}\label{eq:ik} 
         \mathrm{i}_k = y_{k0}v_{0} + y_{kk}v_k + \sum_{\substack{m \in \mathcal{N}_i \\ m \neq k}} y_{km}v_{m},
    \end{align}
 where $y_{km}$ represents the admittance between two nodes, $k,m \in \mathcal{N}_i$, $y_{k0}$ is the admittance between the node $k$ and the slack node, and $v_0$ is the voltage in the slack node.  In \eqref{eq:ik}, the first term denotes the relation of all nodes with the slack node, the second term references the relation of each node with itself, and the last term is associated with the relation of each node with other nodes excluding the slack node and itself.  
    It is possible to group the three terms of the right-hand side expression in \eqref{eq:ik} as the summation
     \begin{align}
          \mathrm{i}_k = \sum_{\substack{m \in \mathcal{N}_i}} y_{km}v_m.
          \label{eq:ik1}
    \end{align}
 Therefore, using the current nodal definition in \eqref{eq:ik1}, the nodal power can be written as
    \begin{align} 
          s^*_k & =   \sum_{\substack{m \in \mathcal{N}_i}}{y_{km}v_k^* v_\text{m} },   \label{eq:s} 
    \end{align}
 which is a non-linear expression due to the product of two complex variables. Defining $f(v_k^*,v_{m})=v_k^* v_{m}$ as the non-linear function, in which each term is independent,  we linearize around the points $v_{k0}^*$ and $v_{m0}$ using 1st-degree Taylor polynomial expansion for two variables, and Wirtinger derivatives \cite{Garces_linear_16}. Thus, we obtain the linearized expression
    \begin{align} 
        f(v_k^*,v_{m}) \approx v_{m0}v_k^* + v_{k0}^* v_{m} - v_{k0}^* v_{m0}. 
    \end{align}
 Therefore, by replacing the linearized expression in \eqref{eq:s}, the net complex power can be written as 
    \begin{align} \label{eq:slin}
         s_k^* =  \sum_{\substack{m \in \mathcal{N}_i}}{y_{km}(v_{m0}v_k^* + v_{k0}^* v_{m} - v_{k0}^* v_{m0})},
    \end{align}
 where the nodal impedance $y_{km}$ and the initial voltages $v_{k0}$ values are constants.

  \subsubsection{Unbalanced-intertemporal linear power flow}

  {We extend the formulation of the linear power flow described in the previous subsection to a power flow formulation of unbalanced MGs for hourly operation, which we call unbalanced-intertemporal linear power flow.} The net injected power expressed in \eqref{eq:slin} can be generalized for a microgrid $i \in \mathcal{M}$ a time $t \in \mathcal{T}$ as 
      \begin{align}\label{eq:s3P-lin-i-t}
      \begin{split}
        s_{i,\phi,k}^*(t) &=  \sum_{\substack{m \in \mathcal{N}_i \\ m \neq k}}{y_{i,\phi,km}(v_{i,\phi,m0}v_{i,\phi,k}^*(t)  + v_{i,\phi,k0}^* v_{i,\phi,m}(t)} \\ &- v_{i,\phi,k0}^* v_{i,\phi,m0}), 
     \end{split}
    \end{align}
     in which the currents, powers and voltages per each node $k \in \mathcal{N}_i$ have different values per phase $\phi \in \mathcal{F}$. This expression of linearized power flow can be expanded as 
        \begin{align} \label{eq:s3Plin1}
             &s_{i,\phi,k}^*(t) =\Bigg(\sum_{\substack{m \in \mathcal{N}_i}}{y_{i,\phi,km}v_{i,m0}} \Bigg)v_{i,\phi,k}^*(t) \; + \nonumber  \\
             \sum_{\substack{m \in \mathcal{N}_i}}&{y_{i,\phi,km}v_{i,k0}^* v_{i,\phi,m}(t)} \;
             - \sum_{\substack{m \in \mathcal{N}_i}}{y_{i,\phi,km}v_{i,\phi,k0}^* v_{i,\phi,m0}},
        \end{align}
       and after algebraic manipulations, the obtained matrix representation is
        \begin{align} \label{eq:s3Plin2}
           \qquad & S_{i,3\phi}^*(t)  =  \operatorname{diag}(Y_{i,3\phi} V_{l_i})V_{i,3\phi}^*(t) \; +  \nonumber \\  (\operatorname{diag}&(V_{l_{i,3\phi}}^*)Y_{i,3\phi})V_{i,3\phi}(t) 
              -  \operatorname{diag}(V_{l_{i,3\phi}})(Y_{i,3\phi} V_{l_{i,3\phi}}^*),
        \end{align}
     in which $S_{i,3\phi} \in \mathbb{C}^{3n}$ is a column vector with the three-phase net complex power of all nodes, $Y_{i,3\phi} \in \mathbb{C}^{3n\times 3n}$ is the three-phase admittance matrix of the network, $V_{i,3\phi} \in \mathbb{C}^{3n}$ is the column vector of the three-phase nodal voltages and $V_{l_{i,3\phi}}\in \mathbb{C}^{3n}$ is the column vector of three-phase initial voltage values.  

     In this power flow formulation, the current, voltage, and power vectors contain the three phases $a,b$, and $c$, which are denoted with the symbol $3\phi$, i.e.,
    \begin{align} 
      S_{i,3\phi}(t) &=  \begin{bmatrix} S_{i,a}(t) \hspace{5pt}  S_{i,b}(t) \hspace{5pt} S_{i,c}(t) \end{bmatrix}, \nonumber \\
      V_{i,3\phi}(t) &=  \begin{bmatrix} V_{i,a}(t)  \hspace{5pt} V_{i,b}(t) \hspace{5pt} V_{i,c}(t) \end{bmatrix}, \nonumber  \\ 
      I_{i,3\phi}(t) &=  \begin{bmatrix} I_{i,a}(t)  \hspace{5pt} I_{i,b}(t) \hspace{5pt} I_{i,c}(t)  \end{bmatrix}.
    \end{align}

   Note that the $3\phi$ notation does not refer to a product. It refers to a three-phase vector and matrices.  
   
   Three-phase unbalanced admittance matrix, $Y_{i,3\phi}$, is obtained through the Kronecker product between the identity matrix and the balanced admittance matrix, {$Y_{i}$,} as 
    \begin{align}\label{eq:Y3P}
         Y_{i,3\phi} = \mathbb{I}_3 \otimes Y_{i}.  
    \end{align}
    
  {The variable $Y_{i}$ is the result of a quadratic expression given by $A_i^T Y_{pi} A_i$, where $A_i \in \mathbb{R}^{\mathcal{N}_ix\mathcal{E}_i}$ is the incidence matrix of the graph $\mathcal{G}_i$ of the microgrid $i \in \mathcal{M}$.  And, $Y_{pi}$ is a diagonal matrix with the series admittance of each branch}.

     Accordingly, it is possible to define an unbalanced  three-phase grid-connected microgrid represented by an affine space given by
     \begin{align} \label{eq:MF1-U}
            S_{i,3\phi}^*(t) = R_{i,3\phi} V_{i,3\phi}^*(t) + U_{i,3\phi} V_{i,3\phi}(t) + Z_{i,3\phi},
        \end{align}
    with $R_{i,3\phi},U_{i,3\phi} \in \mathbb{C}^{3n\times 3n}$ and $Z_{i,3\phi} \in \mathbb{C}^{3n}$ constant matrices defined as follows
        \begin{align}
            R_{i,3\phi} &=  \operatorname{diag}(Y_{i,3\phi} V_{l_{i,3\phi}}), \nonumber \\
            U_{i,3\phi} &= \operatorname{diag}(V_{l_{i,3\phi}}^*)Y_{i,3\phi}, \nonumber \\
            Z_{i,3\phi} &= -\operatorname{diag}(V_{l_{i,3\phi}})(Y_{i,3\phi} V_{l_{i,3\phi}}^*). 
        \end{align}  

  \subsubsection{Power flow constraints}

    Power flow constraints are described in the set  $\mathcal{X}_i^1(t)$, which has the dynamic of the flows of each simulated MG.  In this case, the variables are the complex nodal power and voltages per phase and time as
   \begin{multline}
        \mathcal{X}_i^1(t) = \big\{ S_{i,3\phi}(t), \; V_{i,3\phi}(t) \in \mathbb{C}^{3n} \; | \\ S_{i,3\phi}^*(t) - R_{i,3\phi} V_{i,3\phi}^*(t) - U_{i,3\phi} V_{i,3\phi}(t) = Z_{i,3\phi} \big\}.
   \end{multline}

  In our model, the first node of a MG is the Point of Common Coupling (PCC), which means PCC is located in the node $k=0$ where constraints fix the voltage $\mathcal{X}_i^2(t)$ as follows
        \begin{multline}
            \mathcal{X}_i^2(t) = \Bigg\{ v_{i,a,0}, v_{i,b,0}, v_{i,c,0} \in \mathbb{C} \; | \; v_{i,a,0} = 1 + 0j, \\   v_{i,b,0} = - \frac{1}{2} - \sqrt{\frac{3}{2}}j,  \quad v_{i,c,0} = - \frac{1}{2} + \sqrt{\frac{3}{2}}j  \Bigg\}.
        \end{multline}
 In addition, the reactive power at PCC $p_{i,pcc}(t) \in \mathbb{R}$ corresponds to the sum of the real part of the power at the first node per phase and the complex power vector $S_{i,3\phi}(t)$, which can be expressed as
     \begin{multline}
            \mathcal{X}^3_i(t) = \{  s_{i,a,0}(t), \; s_{i,b,0}(t), \; s_{i,c,0}(t) \in \mathbb{C}, \\ s_{i,pcc}(t) \in \mathbb{C},  p_{i,pcc}(t) \in \mathbb{R} \quad | \\  s_{i,pcc}(t) - s_{i,a,0}(t) - s_{i,b,0}(t) - s_{i,c,0}(t) = 0, \\
            p_{i,pcc}(t) - \operatorname{real}(s_{i,pcc}(t)) = 0  \}.
        \end{multline}

 \subsection{Microgrid power balance}
   The net injected power vector, $S_{i,3\phi}(t) \in \mathbb{C}^{3n}$, is a variable of the unbalanced power flow for all nodes of the microgrid $i \in \mathcal{M}$ at all phases $\phi \in \mathcal{F}$ and time $t \in \mathcal{T}$.  Moreover, $S_{i,3\phi}(t)$ is the difference between generation and demand power expressed as
      \begin{align}\label{eq:cs}
            S_{i,3\phi}(t) = S^G_{i,3\phi}(t) - S^D_{i,3\phi}(t),
        \end{align}
  where $ S^G_{i,3\phi}(t)$, $S^D_{i,3\phi}(t) \in \mathbb{C}^{3n}$ are the generation and demand power vectors respectively.

  The generation power vector, i.e., $S^G_{i,3\phi}(t)$, in the MG $i \in \mathcal{M}$ at all phases $\phi \in \mathcal{F}$ and time $t \in \mathcal{T}$, is the sum of PV generation and discharge power of the batteries, as shown next 
    \begin{align} 
         S^G_{i,3\phi}(t) = S^{pv}_{i,3\phi}(t) + S^{bd}_{i,3\phi}(t),
     \end{align}
    where $S^{pv}_{i,3\phi}(t)$, $S^{bd}_{i,3\phi}(t) \in \mathbb{C}^{3n}$, in which the value of power generation in the nodes with no DER is zero.  

    The demand power $S^D_{i,3\phi}(t) \in \mathbb{C}^{3n}$ is given by  
     \begin{align}
         S^D_{i,3\phi}(t) = S^L_{i,3\phi}(t) + S^{bc}_{i,3\phi}(t),
     \end{align}
     where $S^L_{i,3\phi}(t)$, $S^{bc}_{i,3\phi}(t) \in \mathbb{C}^{3n}$ are vectors of load power and batteries charge power respectively.

     Thus, the power balance is written as the following constraint set,
        \begin{multline}
            \mathcal{X}_i^4 = \big\{S_{i,3\phi}(t), S^{pv}_{i,3\phi}(t), S^{bd}_{i,3\phi}(t), S^{bc}_{i,3\phi}(t) \in \mathbb{C}^{3n} \; | \\ 
            - S_{i,3\phi}(t) + S^{pv}_{i,3\phi}(t) + S^{bd}_{i,3\phi}(t) - S^{bc}_{i,3\phi}(t) = S^L_{i,3\phi}(t) \big\},
        \end{multline}
    where the only parameter is the load power vector \mbox{$S^L_{i,3\phi}(t) \in \mathbb{C}^{3n}$}.

 \subsection{Distributed energy resources (DER)} 
  We consider two types of DER: PV systems and Batteries Energy Storage Systems (BESS), with the capacity to give or absorb reactive power.  

   \subsubsection{Photovoltaic systems (PV)}
     Photovoltaic systems are the distributed generation in the microgrids model.  The maximum power of each PV system, i.e., $ \overline{p}^{pv}_{i,\phi,k}(t)$  of the MG $i \in \mathcal{M}$ at node $k \in \mathcal{N}_i$ time $t \in \mathcal{T}$, is modeled as 
        \begin{align} 
            \overline{p}^{pv}_{i,\phi,k}(t) = p^{mpp}_{i,\phi,k} B^{ir}_i  \varphi^{ir}_i(t)  C^{tem}_{i,\phi,k}(\tau(t)) \eta^{inv}_{i,\phi,k}(t),
            \label{eq:PV_max-3p}
        \end{align}
     where $ p^{ppm}_{i,\phi,k}$ is the rated power at the maximum point of the power of the PV array in kW, for $1 {kW}/{m^2}$ of irradiance.  $B^{ir}_{i,\phi}$ is the base irradiance with a value of 0.8 ${kW}/{m^2}$, $\varphi_{i,\phi}^{ir}(t)$ is the value of the duty irradiance curve at time $t$, and $C_{i,\phi,k}^{tem}(\tau(t))$ is the value of the correction factor due to the temperature curve value $\tau(t)$ at time $t \in \mathcal{T}$. $\eta^{inv}_{i,\phi,k}(t)$ is the inverter efficiency curve. Thus, the maximum PV power, $\overline{p}^{pv}_{i,\phi,k}(t)$, is a function that depends of temperature, $\tau(t)$, and inverter efficiency, $\eta^{inv}_{i,\phi,k}(t)$ and time $t \in \mathcal{T}$ at node $k \in \mathcal{N}_i$ at phase $\phi \in \mathcal{F}$ \cite{OpenDSS_PV-Inv_19, OpenDSS_Manual_18}.

     To include the PV generation into the power flow formulation, the PV powers of each node $k \in \mathcal{N}_i$ are grouped in a column vector called $P^{pv}_{i,3\phi}(t) \in \mathbb{R}$ which is the real part of the variable $S^{pv}_{i,3\phi}(t)$ at time $t \in \mathcal{T}$. $P^{pv}_{i,3\phi}(t)$ is constrained with its minimum and maximum limits as 
        \begin{align}\label{eq:pv-lim-U}
           0 \leq P^{pv}_{i,3\phi}(t) \leq \overline{P}^{pv}_{i,3\phi}(t).
        \end{align}
    Inequalities should be understood component-wise.
    In case where there are no PV systems at node $k \in \mathcal{N}_i$ and phase $\phi \in \mathcal{F}$, the value of the power in the vector  $p^{pv}_{i,\phi}(t)$ is zero.  

    In general, complex power is the sum of active power and reactive power.  Thus, the PV system's complex power is given by
    \begin{align}
        s^{pv}_{i,\phi,k}=p^{pv}_{i,\phi,k}+jq^{pv}_{i,\phi,k},
    \end{align}
    where $p^{pv}_{i,\phi,k}$ is the active power and the real part of the complex power, and $q^{pv}_{i,\phi,k}$ is the reactive power and the imaginary part of the complex power, both at node $k \in \mathcal{N}_i$, phase $\phi \in \mathcal{F}$, and time $t \in \mathcal{T}$.  Moreover, the apparent power is defined as the magnitude of the complex power. Thus, the apparent power of the PV systems, i.e., $|s^{pv}_{i,\phi,k}(t)|$, can be written as
     \begin{align}
        \big |s^{pv}_{i,\phi,k} (t) \big|=\sqrt{\big(p^{pv}_{i,\phi,k}(t)\big)^2+\big(q^{pv}_{i,\phi,k}(t)\big)^2}.
     \end{align}

     In our model, the apparent power of the PV systems is limited by the maximum PV power $\overline{p}^{pv}_{i,\phi,k}(t)$ as
         \begin{align} 
            \big|s^{pv}_{i,\phi,k}(t) \big| \leq \overline{p}^{pv}_{i,\phi,k}(t).
        \end{align}

    The constraints set of PV systems are expressed next
    \begin{multline}
        \mathcal{X}_i^5 = \big\{ P^{pv}_{i,3\phi}(t) \in \mathbb{R}^{3n}, \; s^{pv}_{i,\phi,k}(t) \in \mathbb{C} \; | \\
         P^{pv}_{i,3\phi}(t) = \operatorname{real} \big(S^{pv}_{i,3\phi}(t)\big), \quad 0 \leq P^{pv}_{i,3\phi}(t) \leq \overline{P}^{pv}_{i,3\phi}(t), \\   \big|s^{pv}_{i,\phi,k}(t)\big| \leq \overline{p}^{pv}_{i,\phi,k}(t) \big\}.
    \end{multline}

   \subsubsection{Storage systems}
   We consider batteries as storage systems. We model the dynamics of the batteries' energy through the State of Energy (SoE). SoE is calculated based on the charge/discharge active power of the batteries. 

   Accordingly, \mbox{$SoE_{i,3\phi}(t) \in \mathbb{R}^{3n}$} is the SoE vector of a microgrid $i \in \mathcal{M}$ at node $k \in \mathcal{N}_i$, phase $\phi \in \mathcal{F}$, and time $t \in \mathcal{T}$, and it is modeled as 
        \begin{multline}
            SoE_{i,3\phi}(t+1) = SoE_{i,3\phi}(t) \; +  \\ \Big(\eta^{bc}_{i,3\phi} P^{bc}_{i,3\phi}(t) \; - \; \frac{P_{i,3\phi}^{bd}(t)}{\eta_{i,3\phi}^{bd}} \Big) \Delta t,  
        \end{multline}
   where $P^{bc}_{i,3\phi} \in \mathbb{R}^{3n}$ is the batteries charge power vector and $P^{bd}_{i,3\phi} \in \mathbb{R}^{3n}$ is the discharge power vector. Similarly to the PV power vector, the value of the charge/discharge power and SoE is zero in nodes without batteries.  

  Additionally, the continuity of SoE between the last and first hour needs to be guaranteed, i.e.,
        \begin{align}
            \begin{split}
            SoE_{i,3\phi}(t_{\text{last}}) =  SoE_{i,3\phi}(t_\text{ini}), 
            \end{split} 
        \end{align}
     where $t_\text{ini} $ is the first element of the set $\mathcal{T}$ and $t_\text{last} $ is the last element.

    On the other hand, SoE and charge/discharge power are limited by the maximum capacities of the batteries by the following box constraints
        \begin{align}
            (1-DoD_{i,3\phi})\overline{SoE}_{i,3\phi} &\leq SoE_{i,3\phi}(t) \leq \overline{SoE}_{i,3\phi} \\
             0 &\leq P^{bc}_{i,3\phi}(t) \leq \overline{P}^{b}_{i,3\phi}, \\
             0 & \leq P^{bd}_{i,3\phi}(t) \leq \overline{P}^{b}_{i,3\phi},
        \end{align}
    where $\overline{SoE}$ is the maximum SoE capacity given by the physical battery specifications.  The inferior limit of SoE is determined by the Depth of Discharge (DoD) with the purpose of not fully discharging the batteries. The superior SoE limit is bounded by $\overline{SoE}$.

   Power flow is linked with the batteries model through charge and discharge complex power.  The active power is the real part of the complex power.  Thus, we can write the charge active power as $ P^{bc}_{i,3\phi}(t) = \operatorname{real}\big(S^{bc}_{i,3\phi}(t)\big)$ and discharge active power as $P^{bd}_{i,3\phi}(t) = \operatorname{real}\big(S^{bd}_{i,3\phi}(t)\big)$. 

    In addition, we introduce constraints of charge and discharge apparent power to restrict the reactive power as follows
        \begin{align} 
           \big |s^{bc}_{i,\phi,k}(t)\big| \leq \overline{p}^{b}_{i,\phi,k}, \quad \big |s^{bd}_{i,\phi,k}(t) \big| \leq \overline{p}^{b}_{i,\phi,k}.
        \end{align} 

    Finally, the batteries constraints can be grouped as 
   \begin{multline}
      \mathcal{X}_i^6(t) = \big\{SoE_{i,3\phi}(t), P^{bc}_{i,3\phi}(t), P^{bd}_{i,3\phi}(t), S^{bc}_{i,3\phi}(t), \\ S^{bd}_{i,3\phi}(t) \in \mathbb{R}^{3n} \; | \;
                       - SoE_{i,3\phi}(t+1) + SoE_{i,3\phi}(t) \\
                       + \; \Big(\eta^{bc}_{i,3\phi} P^{bc}_{i,3\phi}(t)  \; - \; \frac{P_{i,3\phi}^{bd}(t)}{\eta_{i,3\phi}^{bd}} \Big) = 0, \\ SoE_{i,3\phi}(t_{\text{last}}) - SoE_{i,3\phi}(t_0) = 0, \\ (1-DoD_{i,3\phi})\overline{SoE}_{i,3\phi} \leq SoE_{i,3\phi}(t) \leq \overline{SoE}_{i,3\phi} \\
             0 \leq P^{bc}_{i,3\phi}(t) \leq \overline{P^{b}}_{i,3\phi}, \; 0 \leq P^{bd}_{i,3\phi}(t) \leq \overline{P^{b}}_{i,3\phi} \\ 
             P^{bc}_{i,3\phi}(t) = \operatorname{real}\left(S^{bc}_{i,3\phi}(t)\right), P^{bd}_{i,3\phi}(t) = \operatorname{real}\left(S^{bd}_{i,3\phi}(t)\right), \\  |s^{bc}_{i,\phi,k}(t)| \leq \overline{p}^{b}_{i,\phi,k}, \; |s^{bd}_{i,\phi,k}(t)| \leq \overline{p}^{b}_{i,\phi,k} \big\}.
  \end{multline}
  In summary, $\mathcal{X}_i^6(t)$ this set has three components: 1) dynamics and continuity of SoE, 2) box constraints of SoE and charge/discharge active power maximum capacities, and 3) relation between charge/discharge complex and real power.

\subsection{Power quality constraints} \label{s:PQC}

    \subsubsection{Harmonic power flow}
     Fundamental nodal current $I_{i,3\phi}(t) \in \mathbb{C}^{3n}$ of an unbalanced microgrid $i \in \mathcal{M}$ at time $t \in \mathcal{T}$ can be written as 
    \begin{align}
        I_{i,3\phi}(t) = Y_{i,3\phi} V_{i,3\phi}(t), \label{eq:I1}
    \end{align}
    where the vector $V_{i,3\phi}(t)$ is a column vector of the fundamental nodal voltages and  $Y_{i,3\phi}$ is the fundamental admittance matrix.

    The harmonic current $I_{i,3\phi,h}(t) \in \mathbb{C}^{3n}$ can be defined as the product of fundamental current and harmonic spectrum $\psi \in \mathbb{N}$ as 
        \begin{align} 
            I_{i,h,3\phi}(t) = \psi_{i,h,3\phi} I_{i,3\phi}(t),  \label{eq:Ih}
        \end{align}
     for a microgrid $i \in \mathcal{M}$  at time $t \in \mathcal{T}$, and harmonic $h \in \mathcal{H}$ with $\mathcal{H}$ being the set of harmonic frequencies \cite{Castellanos_21}.

    Furthermore, the harmonic current is also defined as the product of the admittance matrix and nodal voltages vector as 
        \begin{align}
            I_{i,h,3\phi}(t) = Y_{i,h,3\phi} V_{i,h,3\phi}(t), \label{eq:Ih1}
        \end{align}
    where $V_{i,h,3\phi}(t) \in \mathbb{C}^{3n}$ is a column vector of the harmonic nodal voltages. $Y_{i,h,3\phi} \in \mathbb{C}^{3n \times 3n}$ is the harmonic admittance matrix which is symmetric and invertible \cite{LectureNotes_Low21}.    $\psi_{i,h,3\phi}$ is the harmonic spectrum, all at the harmonic frequency $h$.  

    Thus, from \eqref{eq:Ih} and \eqref{eq:Ih1}, we obtain the following expression
       \begin{align}
            Y_{i,h,3\phi} V_{i,h,3\phi}(t) &= \psi_{i,h,3\phi} Y_{i,3\phi} V_{i,3\phi}(t). 
        \end{align}

  Finally, we write the harmonic power flow as the constraint set
  \begin{multline}
       \mathcal{X}^7_i(t) = \Big\{ V_{i,h,3\phi}(t), V_{i,3\phi}(t)  \in \mathbb{C}^{3n} \; | \; \\ Y_{i,h,3\phi} V_{i,h,3\phi}(t) - \psi_{i,h,3\phi} Y_{i,3\phi} V_{i,3\phi}(t) = 0   \Big \},
   \end{multline}
    where the set variables are the complex nodal voltages, $V_{i,3\phi}(t) $, and the harmonic nodal voltages, $V_{i,h,3\phi}(t) $, for MG $i \in \mathcal{M}$, node $k \in \mathcal{N}_i$, and time $t \in \mathcal{T}$.

    \subsubsection{Total Harmonic Distortion (THD)}

     According to the IEEE Standard 519 \cite{IEEE519-1992, IEEE519-2014}, THD is a metric of the amount of signal distortion.  THD is the ratio between the root mean square (RMS) value of the harmonic components and the RMS value of the fundamental component.  THD is calculated as
    \begin{align} 
         \text{THD}=\dfrac{\sqrt{\sum\limits_{h \in \mathcal{H}}\mathscr{U}^2_h}}{\mathscr{U}_1} \times 100 \%,
    \end{align}
    where $\mathscr{U}_1$ represents the fundamental signal of voltage or current and $\mathscr{U}_h$ the harmonic voltage or current at harmonic frequency $h$.

    In our formulation, the THD constraint can be written as  
    \begin{align}
        \dfrac{\norm{V_{i\phi,k}(t)}}{|v_{i,\phi,k}(t)|} \leq \overline{\text{THD}}, \label{eq:thd_const}
    \end{align}
    where $\overline{\text{THD}}$ is the upper bound of THD.  The limit allowed of THD for low voltage networks in the PCC is $8\%$, following the IEEE Standard \mbox{519-2014 \cite{IEEE519-2014}}. 

   We linearize the magnitude of the complex voltage around the operating points, i.e., $\operatorname{e}^{\text{j}\theta}$ and $\operatorname{e}^{-\text{j}\theta}$, with phase angles $\theta = 0^\circ and \pm 120^\circ $.  The phase angle $\theta = 0^{\circ}$ represents $\phi = a$, while $\theta = \pm 120^{\circ}$ represent phases $\phi =b$ and $\phi =c$.

    The nodal voltage $v_k$ at node $k \in \mathcal{N}_i$ can be written as $|v_k|\operatorname{e}^{\text{j}\theta}, $ where $|v_k|$ is the magnitude of the voltage, and $\theta$ is the phase angle.  We define $f(v_k, v_k^*) = \sqrt{v_k v_k^*}$ and linearize around the point $ \big(\operatorname{e}^{\text{j}\theta}, \operatorname{e}^{-\text{j}\theta}\big)$.  We use 1st-degree Taylor polynomial expansion in two variables and Wirtinger derivatives over the complex voltage \cite{Wirtinger_27,Garces_18}, obtaining the linearized expression of the voltage magnitude as
    \begin{align}
         |\tilde{v}_k| =  \dfrac{1}{2} \operatorname{e}^{-\text{j}\theta} v_k  + \dfrac{1}{2} \operatorname{e}^{\text{j}\theta} v_k^*, 
    \end{align}
    which is equivalent to
    \begin{align}
         |\tilde{v}_k| = \cos \theta \operatorname{Re}(v_k) + \sin \theta \operatorname{Im}(v_k).
    \end{align}

    Finally, we can write the THD constraint as 
    \begin{multline}
       \norm{V_{i\phi,k}(t)} \; \leq \\ 
       \big(\cos\theta \operatorname{Re}(v_{i,\phi,k}(t)) + \sin\theta \operatorname{Im}(v_{i,\phi,k}(t))\big) \overline{\text{THD}}.
    \end{multline}

    \subsubsection{Voltage limits}

     According to the IEEE Standard 1250-2018 \cite{IEEE1250-2018}, the maximum allowed voltage fluctuation is $\delta = 5\%$.  Thus, we use the surrogate constraint
    \begin{align} 
        \left|{v_{i,\phi,k}(t)-v^{nom}_{i,\phi,k}} \right| \leq \delta v^{nom}_{i,\phi,k},
    \end{align}
    which is a bound on the difference between the nodal voltages and nominal voltage of MG $i \in \mathcal{M}$ at node $k \in \mathcal{N}_i, $ phase $\phi \in \mathcal{F}$ and time $t \in \mathcal{T}$. 

   Finally, the THD and voltage limit are grouped in a constraint set as
   \begin{align}
       \mathcal{X}^8_i(t) = &\Big\{ V_{i,\phi,k}(t) \in \mathbb{C}^{h\times3n}, v_{i,\phi,k}(t) \in \mathbb{C} \; | \nonumber \\
       &\norm{V_{i,\phi,k}(t)} -  \big(\cos\theta_{\phi} \operatorname{Re}(v_{i,\phi,k}(t))  \nonumber \\ 
       &+ \sin\theta_{\phi} \operatorname{Im}(v_{i,\phi,k}(t))\big) \overline{\text{THD}}_{i} \leq 0, \nonumber \\
       &\big|{v_{i,\phi,k}(t)-v^{nom}_{i,\phi,k}}\big| \leq \delta v^{nom}_{i,\phi,k}  \Big \}. 
   \end{align}

  \subsection{Interaction constraints}
    We define the active power at PCC as $p_{i,pcc}(t) \in \mathbb{R}$ of MG $i \in \mathcal{M}$ in the set $\mathcal{X}^3_i(t)$. Positive values of $p_{i,pcc}(t)$, at time $t \in \mathcal{T}$, means MG $i$ needs external sources to supply its demand. In contrast, negative values of $p_{i,pcc}(t)$ indicate a generation surplus by MG $i$, which can be sold to other MGs or delivered to the DSO.  

  Active power exchanges, with other agents $j \in \mathcal{A}$, take place at the PCC. In this paper, active power exchanges are referred to as interactions.  In other words, MG $i \in \mathcal{M}$ can interact with other agents $j$ selling or purchasing energy depending on its demand or surplus power at the PCC. Thus, we can write the interactions constraints as follows
      \begin{multline}
             \mathcal{X}^9_i(t) = \Bigg \{ \mathscr{P}_{ji}(t), \mathscr{P}_{e}(t) \in \mathbb{C} \; | \\  \; p_{i,pcc}(t) = \sum_{\substack{e \in \mathcal{E} \\ e \notin (i,i)}} {\mathscr{P}_{e}(t) },  
              \sum_{\substack{j \in \mathcal{M} \\ i \neq j}} {\mathscr{P}_{ij}(t) } \leq \operatorname{max} \{ \Hat{\mathscr{P}}_{i0}(t),0 \}, \\ 
               \quad \sum_{\substack{j \in \mathcal{M} \\ i \neq j}} {\mathscr{P}_{ij}(t) } \leq |\operatorname{min} \{ \Hat{\mathscr{P}}_{i0}(t),0 \}|, \quad \mathscr{P}_{ij} \geq 0 \Bigg\}  
        \end{multline}
 where $\mathscr{P}_{ij}(t)$ is the energy sold from MG $i$ to agent~$j$.  A MG can only have one role at a time $t \in \mathcal{T}$, which means a MG can not purchase or sell energy at the same time instant. We impose this behavior with the constraint $\mathscr{P}_{ij}(t) \times \mathscr{P}_{ji}(t) = 0$.


 \section{Energy management system in a Local Energy Market (LEM)} \label{S3-OPTmodel}

  This section will define the {proposed } LEM and the agents' energy interactions, explaining the dynamic market framework.  
 
  {LEM is a market framework where prosumers and consumers can transact directly and at different electricity prices. To achieve the best balancing between local supply, local demand, and grid exchange, LEMs allocate the local resources at the lowest possible cost \mbox{\cite{Sumper_19}}.}
  
\begin{figure}[htb]  
        \centering
        \includegraphics[width=\linewidth]{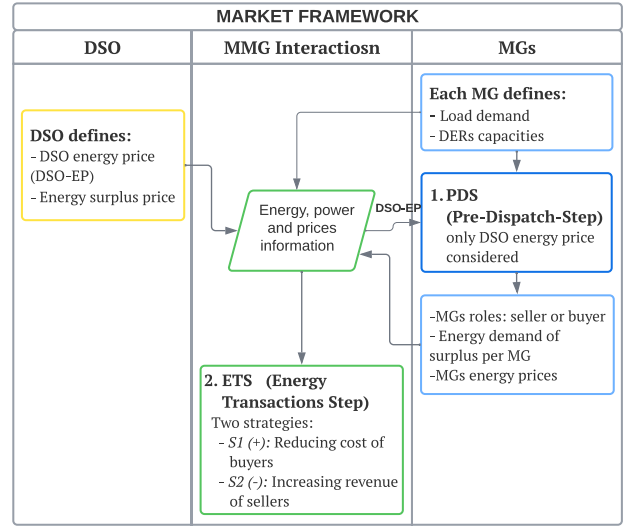}
        \caption{Local market framework.  The interactions are established in a local energy market.  Each agent defines its own prices; only DSO has different prices for selling and purchasing energy.  The structure of the market is divided into two main steps: 1) the Pre-Dispatch Step (PDS) and 2) the Energy Interactions Step (ETS).}
        \label{fig:framework}
    \end{figure}

 Figure \ref{fig:framework} shows the schematic representation of the framework. In this market, no contracts are established. However, transactions among agents are defined the previous night. We propose two steps: Pre-Dispatch (PDS) and Energy Transactions (ETS).  
  
  { Our proposed market model has a central control model with decentralized decision-making. The first stage, PDS, establishes the upper boundaries for energy transactions for each MG. An MG transforms into a buyer when it requires more energy to meet its own demand. Additionally, if an MG has extra energy, it becomes a seller. It is important to remember that these positions may change depending on the circumstances (energy surplus or energy requirement) at each hour of the day.  In PDS, the decision is centralized from MGs because each MG operator sets its own energy prices and power exchange limits. However, from the perspective of the market platform, this choice is decentralized because each MG makes it on its own, i.e., the market does not decide on the costs of the energy MGs or their energy upper bounds.}
  
   {As a result, MGs upper bounds obtained from PDS are becoming "box constraints" for the energy management system of the second stage, i.e., ETS. Accordingly, any MG with a selling position has a window of time to satisfy other MGs' needs based on MGs prices. A market operator carries out the EMS defined in the ETS, which is based on two strategies: S1 and S2. S1 refers to lowering buyer costs, whereas S2 favors increasing seller revenue.}

\begin{figure}[htb!]
     \centering
     \includegraphics[scale=0.4]{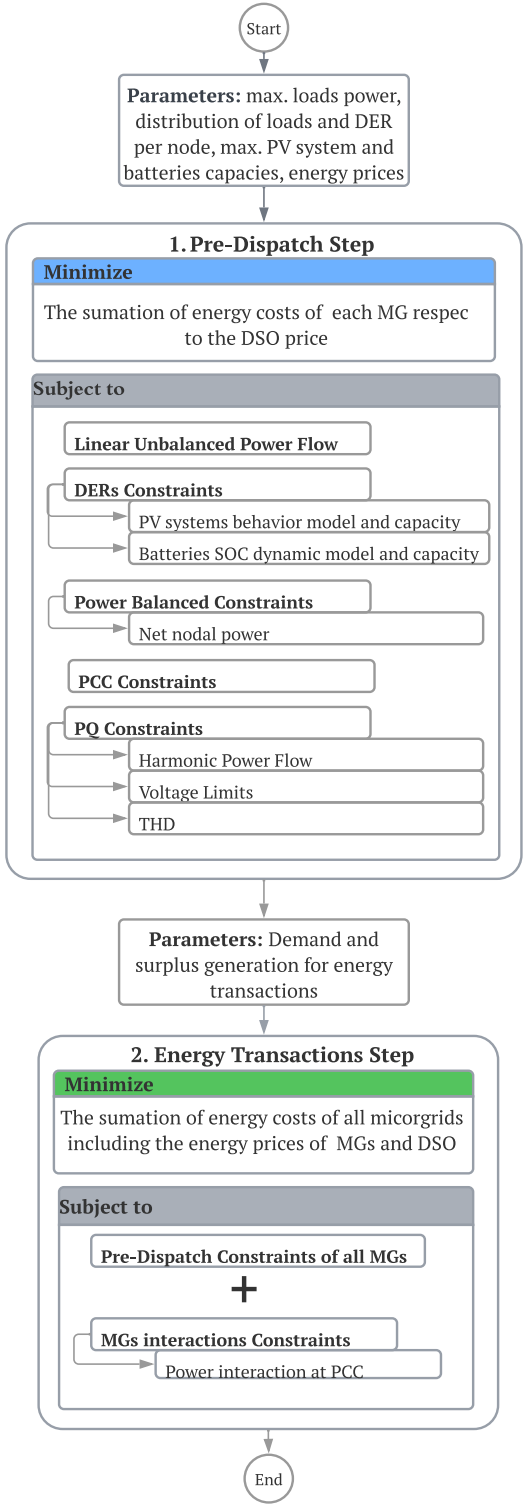}
     \caption{Pre-dispatch (PDS) and energy transactions (ETS) optimization schemes.  PDS combines the MGs' energy management system into a single optimization problem by adding the MGs' objective functions subject to all MGs' operational and PQ constraints.  The result of the PDS is the maximum energy MGs can trade in the market, which is the input of ETS.  ETS includes the interaction constraints of the model. }
     \label{fig:Opt_model}
 \end{figure}

  PDS and ETS are detailed in Figure \ref{fig:Opt_model}, where each step's optimization structure with block constraints is presented.   {Both steps are based on market schemes for the day ahead.}

  The optimization problem of the PDS is 
  \begin{align} 
          {\min_{}} &\quad \sum_{t \in \mathcal{T} }  \zeta_{0}(t)  \widehat{\mathscr{P}}_{i0}(t)  \label{eq:pds} \\ \nonumber
          &\text{ subject to} \\ \nonumber
          &\qquad \mathcal{X}_i^{PD}(t) = \bigcap_{l}^8X_i^{l}(t) \\   \nonumber
          & \qquad \forall i \in \mathcal{M}, \forall t \in \mathcal{T},
    \end{align}
   where the energy cost of the MGs is minimized without interactions among them.  Moreover,  MGs differ in their interactions and parameters, such as loads, DER power capacities, and energy prices. Interactions also happen with the DSO, and the DSO energy and surplus cost is the same.  PDS determines the maximum capacity for energy exchange among MGs. In other words, the maximum amount of energy each MG can sell to others and, at the same time, the energy each MG needs to purchase from the other agents.

  After PDS, the roles of sellers and buyers are established according to the MGs' demand requirements and power sales capacities.  Subsequently, the inputs for the ETS are determined by PDS, which are MGs demand, maximum capacities, and roles.

 The solution of \eqref{eq:pds} is $\widehat{\mathscr{P}}_{i0}^*(t)$.  Positive values of $\widehat{\mathscr{P}}_{i0}^*(t)$ indicate the demand of the MGs, while negative values indicate the surplus generation at time $t \in \mathcal{T}$.  We use the optimal values of $\widehat{\mathscr{P}}_{i0}(t)$ as parameters of the ETS.

  The Energy Transaction Step (ETS) is the complete dispatch formulation that integrates the optimization problems of all MGs.  The proposed model of the ETS is 
 \begin{align}\label{eq:ets}
        {\min_{}} &  \sum_{t \in \mathcal{T} } \Bigg( \zeta_{0}(t) \sum_{i \in \mathcal{M}} { \mathscr{P}_{0i}(t)} -  \widehat{\zeta}_{0} \sum_{i \in \mathcal{M}} \mathscr{P}_{i0}  + \sum_{e \in \mathcal{E}} {\zeta_{e}(t)\mathscr{P}_{e}(t)} \Bigg) \\
        &\text{ subject to} \nonumber \\ \nonumber
        &  \qquad \mathcal{X}_i^{ET}(t) = \bigcap_{l}^9X_i^{l}(t) \\ \nonumber 
        &  \qquad \forall i \in \mathcal{M}, \forall e \in \mathcal{E}, \forall t \in \mathcal{T},
    \end{align}
 where interactions among agents are established, selecting a criterion between MGs costs or revenues. The  constraints are operational, PQ, and interaction sets detailed in Sections~\ref{s:OIC}. 
 
 { Due to the convex nature of the problem, we can guarantee that the proposed model will result in energy transactions at the lowest possible cost.}

\section{Simulation and results} \label{S6-sim&res}

\begin{figure*}[htb]
             \centering
             \includegraphics[scale=0.35]{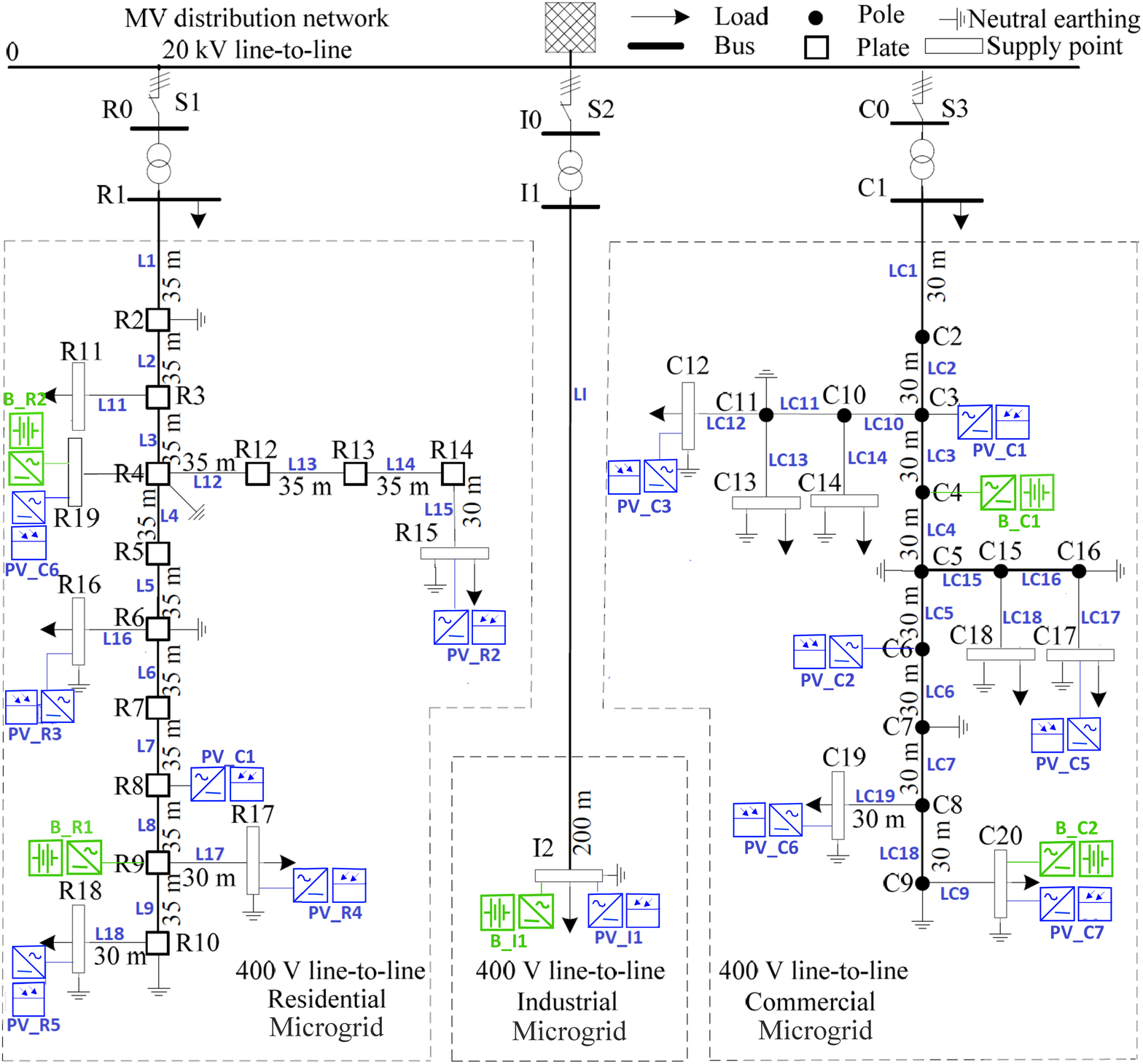}
             \caption{Microgrids topology. The CIGRE European LV distribution network benchmark \cite{CIGRE_2014} has been modified to establish our application case for the proposed model. In our study, each sub-network corresponds to an MG, and we modify the benchmark, including PV generation (boxes in blue) and BESS (boxes in green). Rx represents electric nodes, Lx distribution lines, Bx batteries systems, and PVx PV systems  }
             \label{fig:MGs}
         \end{figure*}

 A test case is proposed based on the CIGRE LV network benchmark with European configuration \cite{CIGRE_2014}.  In our case, the benchmark sub-networks represent the MGs, classified into residential, industrial, and commercial.  Figure~\ref{fig:MGs} shows the system's topology.
 The optimization problem is implemented in CVXPY using the commercial MOSEC solver in Python \cite{Mosek, CVXPY}. 

\subsection{Microgrids parameters} \label{S:param}

 Microgrids are radial, three-phased, and unbalanced with a frequency of 50 Hz and a Medium Voltage (MV) network short circuit power of 100 MVA. The loads are connected along the distribution lines, and the network includes underground and overhead lines. The transformer ratings are 500 kVA, 150 kVA, and 300 kVA for residential, industrial, and commercial MGs.  Moreover, the MGs are connected to the same Point of Common Coupling (PCC).  

 The load apparent power for the balanced case was taken from the CIGRE Benchmark \cite{CIGRE_2014}.   We modified this benchmark to propose unbalanced load distribution displayed in Table \ref{tab:loads}.
 \begin{table}[width=.85\linewidth,cols=4,pos=ht]
    \caption{Maximum loads rating [kVA].}\label{tab:loads}
        \begin{tabular*}{\tblwidth}{@{} LLLLLL@{} }
            \toprule
           Node &  $\boldsymbol{i,k}$ & $\boldsymbol{s_{i,k}^L}$  & $\boldsymbol{s_{i,a,k}^L}$ & $\boldsymbol{s_{i,b,k}^L}$ & $\boldsymbol{s_{i,c,k}^L}$\\
            \midrule
             NR2 & 1,2 & 200 & 60 & 20 & 120  \\
             NR11 & 1,11 & 15 & 10 & 5 & 0 \\
             NR15 & 1,15 & 52 & 17 & 5 & 30  \\
             NR16 & 1,16 & 55 & 35 & 15 & 5  \\
             NR17 & 1,17 & 35 & 5 & 2 & 28  \\
             NR18 & 1,18 & 47 & 2 & 30 & 15  \\
             NI2 & 2,2 & 100 & 35 & 33 & 32  \\
             NC2 & 3,2 & 120 & 20 & 40 & 60  \\
             NC12 & 3,12 & 20 & 5 & 15 & 0  \\
             NC13 & 3,13 & 20 & 0 & 5 & 15  \\
             NC14 & 3,14 & 25 & 20 & 0 & 5  \\
             NC17 & 3,17 & 25 & 1 & 8 & 16  \\
             NC18 & 3,18 & 8 & 0 & 0 & 8  \\
             NC19 & 3,19 & 16 & 5 & 11 & 0  \\
             NC20 & 3,20 & 8 & 8 & 0 & 0  \\
            \bottomrule
        \end{tabular*}
    \end{table}

\begin{figure}[htb]
    	\centering
            \includegraphics[scale=0.55]{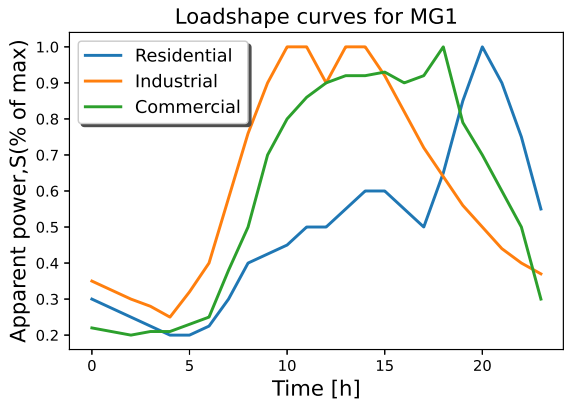}
    	\caption{Daily load profiles of the MGs per hour.  The load shape of the residential MG (blue curve) has its maximum peak around 19h, the industrial MG (orange curve) at noon, and the commercial MG (green curve).  The apparent power is presented as a percentage of the magnitude of the power value, which is applied dynamically to the maximum load parameters per MG.}
    	\label{fig:loadshape}
    \end{figure}

 Figure \ref{fig:loadshape} shows the daily load profiles associated with the residential, industrial and commercial MGs. Moreover, the power factor for the residential, industrial, and commercial MGs is 0.95, 0.85, and 0.9.  

\begin{figure}[htp]
     \centering
     \begin{subfigure}[b]{0.22\textwidth}
         \centering
         \includegraphics[width=\textwidth]{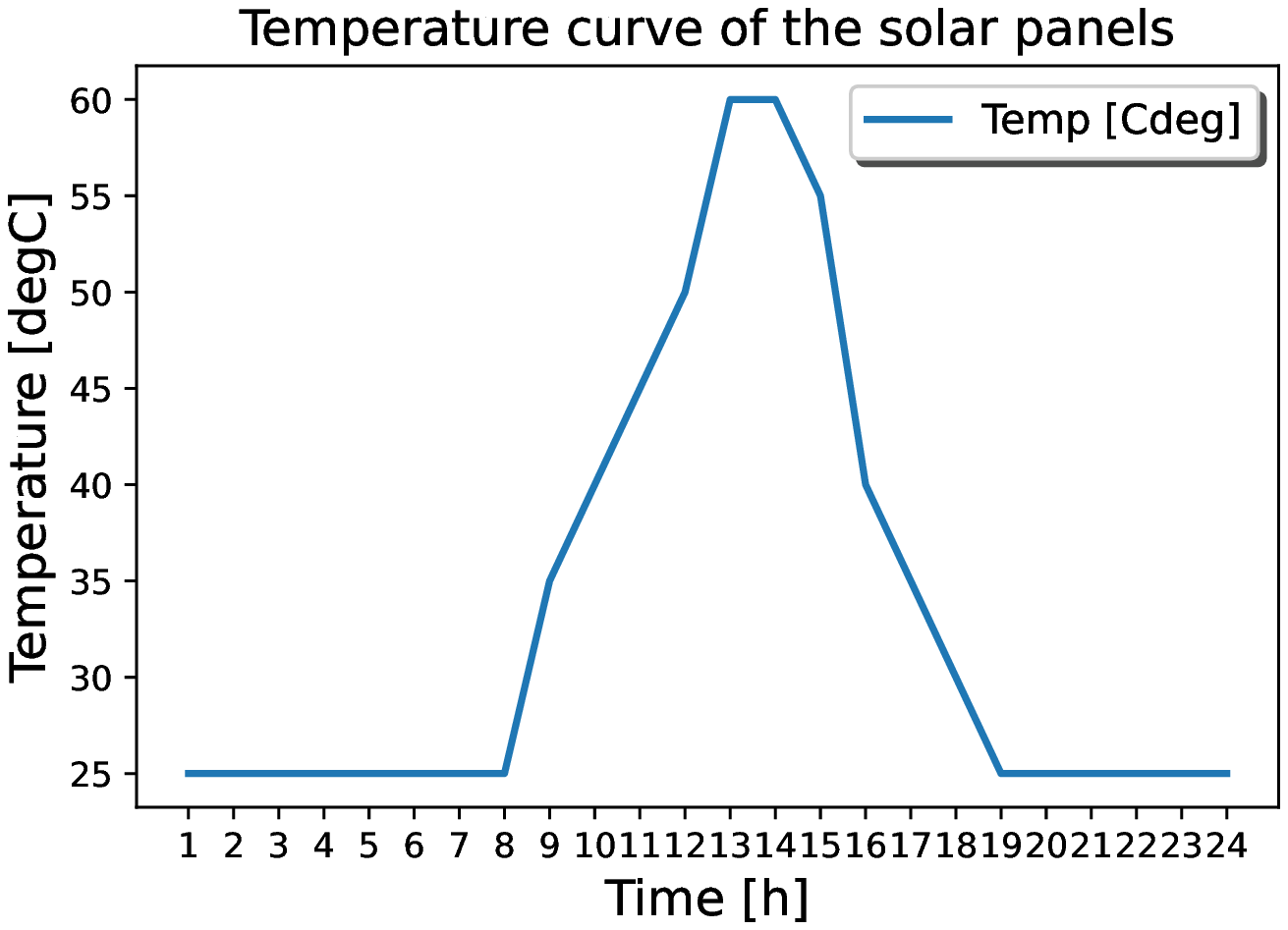}
         \caption{Temperature curve }
         \label{fig:pv-temp}
     \end{subfigure}
     \begin{subfigure}[b]{0.22\textwidth}
         \centering
         \includegraphics[width=\textwidth]{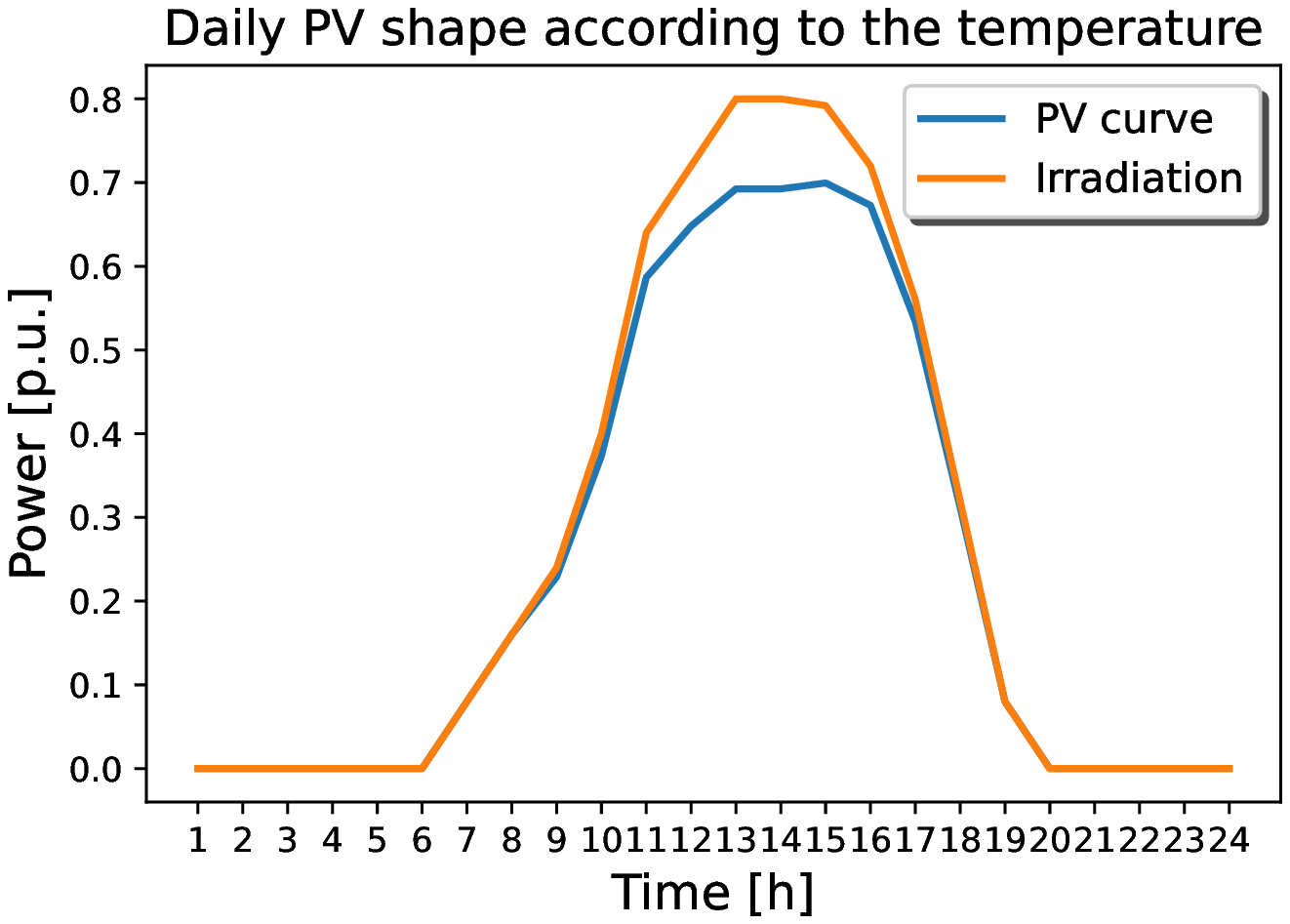}
         \caption{Irradiation and power}
         \label{fig:pv-irrad}
     \end{subfigure}
     \begin{subfigure}[b]{0.24\textwidth}
         \centering
         \includegraphics[width=\textwidth]{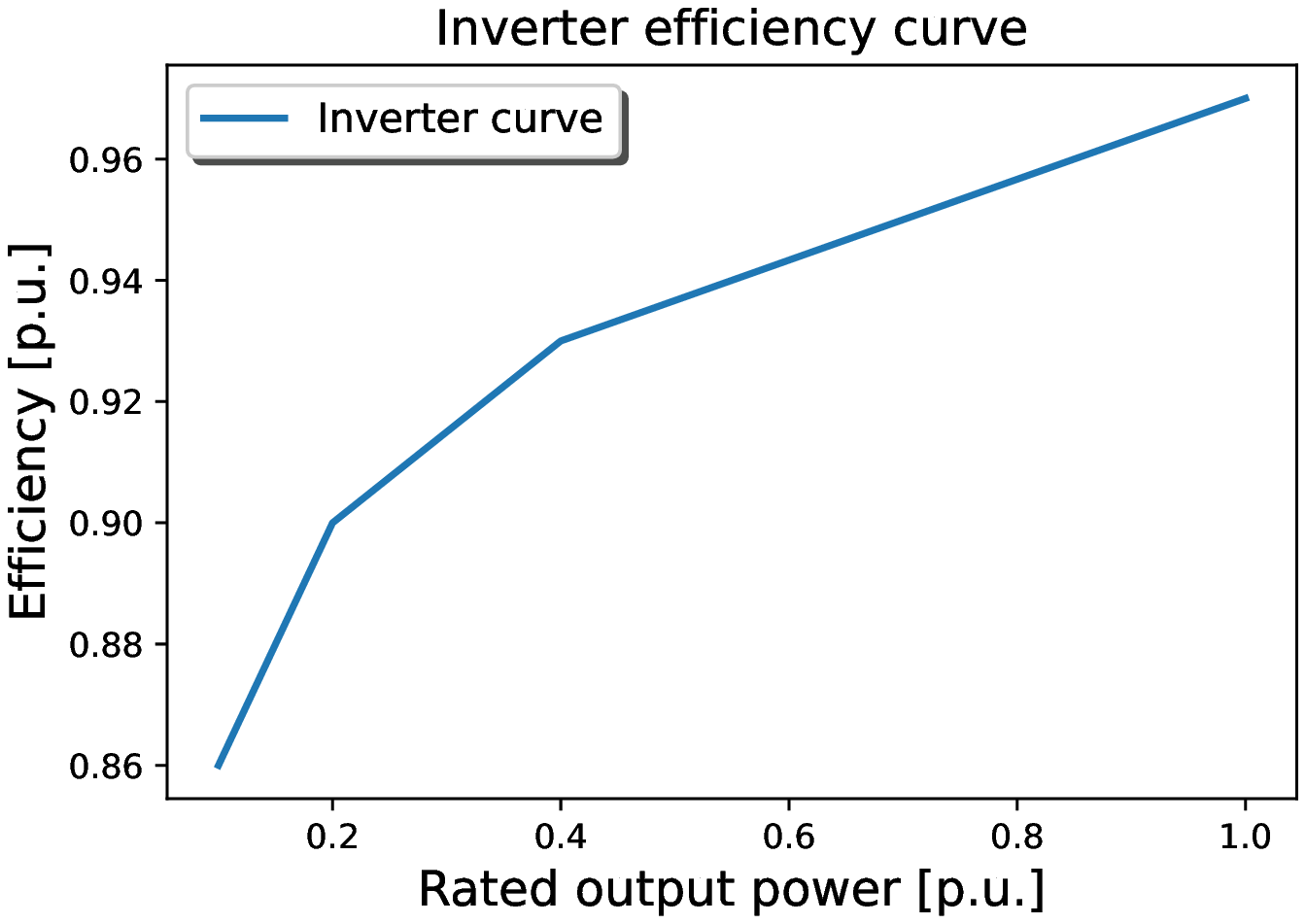}
         \caption{Inverter efficiency curve}
         \label{fig:pv-eff}
     \end{subfigure}
        \caption{Characterization of the PV systems parameters{. (a) shows the solar panels' temperature in Celsius, where the maximum temperature is 60 degrees at noon.  (b) The irradiation curve in orange and the PV curve in blue with the irradiation and temperature effect.  (c) presents the inverter efficiency according to the power in p.u.   } }
        \label{fig:pv-param}
  \end{figure}

 The maximum capacity of the PV systems \eqref{eq:PV_max-3p} includes the effects of irradiance, temperature, and inverter efficiency on the maximum power generated by the PV systems.  The parameters of the PV systems are shown in Figure \ref{fig:pv-param} and the maximum rating of the PV systems are shown in Table \ref{tab:PVs}.
    \begin{table}[width=\linewidth,cols=4,pos=ht]
    \caption{DERs maximum capacities.}\label{tab:PVs}
        \begin{tabular*}{\tblwidth}{@{} LLLLLLLLLL@{} }
            \toprule
           Node & $\boldsymbol{k}$   & $\boldsymbol{p_{i,k}^{ppm}}$ & $\boldsymbol{p_{i,a,k}^{ppm}}$ & $\boldsymbol{p_{i,b,k}^{ppm}}$ & $\boldsymbol{p_{i,c,k}^{ppm}}$ & $\boldsymbol{\overline{soc}_i}$ & $\boldsymbol{\overline{soc}_{i,a,k}}$ & $\boldsymbol{\overline{soc}_{i,b,k}}$ & $\boldsymbol{\overline{soc}_{i,c,k}}$\\
            \midrule
             {NR8 } & 1,8 & 100 & 50 & 50 & 0 & 0 & 0 & 0 & 0 \\
              {NR9 } & 1,9 & 0 & 0 & 0 & 0 & 100 & 50 & 50 & 0  \\
             {NR15 } & 1,15 & 50 & 0 & 25 & 25 & 0 & 0 & 0 & 0  \\
             {NR16 } & 1,16 & 56 & 28 & 0 & 28 & 0 & 0 & 0 & 0  \\
              {NR17 } & 1,17 & 34 & 17 & 0 & 17 & 0 & 0 & 0 & 0  \\
              {NR19 } & 1,19 & 70 & 0 & 35 & 35 & 100 & 0 & 50 & 50  \\
              {NI2 } & 2,2 & 120 & 40 & 40 & 40 & 90 & 30 & 30 & 30  \\
             {NC3 } & 3,3 & 50 & 25 & 0 & 25 & 0 & 0 & 0 & 0  \\
              {NC4 } & 3,4 & 0 & 0 & 0 & 0 & 100 & 50 & 50 & 0  \\
              {NC6 } & 3,6 & 50 & 0 & 25 & 25 & 0 & 0 & 0 & 0  \\
              {NC12 } & 3,12 & 70 & 35 & 35 & 0 & 0 & 0 & 0 & 0  \\
              {NC17 } & 3,17 & 30 & 0 & 15 & 15 & 0 & 0 & 0 & 0  \\
             {NC19 } & 3,19 & 15 & 0 & 7.5 & 7.5 & 0 & 0 & 0 & 0  \\
             {NC20 } & 3,20 & 10 & 0 & 5 & 5 & 50 & 25 & 0 & 25  \\
            \bottomrule
        \end{tabular*}
    \end{table}

  The location of the PV systems of residential MG was inspired from \cite{CIGRE_net}, where PV generators and batteries replaced DER, and some DER node positions were changed.  We selected the PV systems' location for industrial and commercial MGs, considering the network's endpoints.

  The simulation includes a harmonic spectrum in the nodes of the PV system, which contains six odd harmonics from the harmonic frequency of three to thirteen, which are given by the vector
  \begin{align*}
      \psi = [0.088,  2.215,  0.754,  0.038,  0.113, 0.0497],
  \end{align*}
  in percentage. The spectrum was obtained from real measurements taken from a proprietary PV system in a microgrid in the Electric, Electronic, and Telmatecommunication Engineering Building at Universidad Industrial de Santander in Bucaramanga, Colombia [7°08'31.1"N 73°07'16.6"W].  Datasets are available by request. 

  \subsection{PDS simulation}

   {The proposed local market framework, shown in Figure~\ref{fig:framework}, has as first step the pre-dispatch step (PDS) optimization model \eqref{eq:pds} which is developed for each MG$i \in \mathcal{M}$.  Each agent's PDS, i.e., MGs, uses the energy sale price from the DSO agent as an input. Along with the specifications and maximum generation capacity of the PV systems, they also need to take into account the battery energy capacities, load profiles, maximum rates, and harmonic spectrum of the generation units.   As a result, each agent MG $i \in mathcal{M}$ manages its own energy, minimizing the operational costs.   Since each MG in our model has zero generation costs, MGs only take the cost of the DSO into account  when their energy resources are not enough to supply the internal demand.}
    
\begin{figure}[htb]
    	\centering
            \includegraphics[scale=0.55]{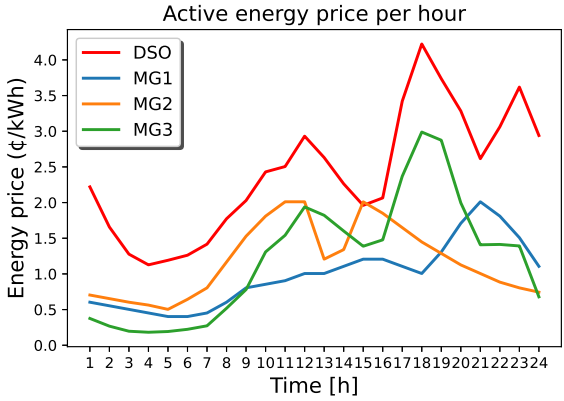}
    	\caption{Energy prices of the agents. The DSO energy price (from DSO to MGs) $\zeta^{\epsilon}_0$ in the red line is the highest energy price in the market. The dashed line's surplus energy (from MGs to DSO) is three times less than the DSO energy price. MG$_1$, MG$_2$, and MG$_3$ are denoted by blue, orange, and green colors.  }
    	\label{fig:prices}
    \end{figure}

   {The PDS establish the DSO energy price $\zeta_0$ equal to the surplus energy $\Hat{\zeta}_0$. The sale price of the DSO agent per time is shown in Figure \ref{fig:prices}.  The prices were taken from UK day-ahead energy spot market data from European spot market (epexspot) in December of 2016 \cite{EpexSpot}.}
    
\begin{figure}[htb]
    	\centering
            \includegraphics[scale=0.5]{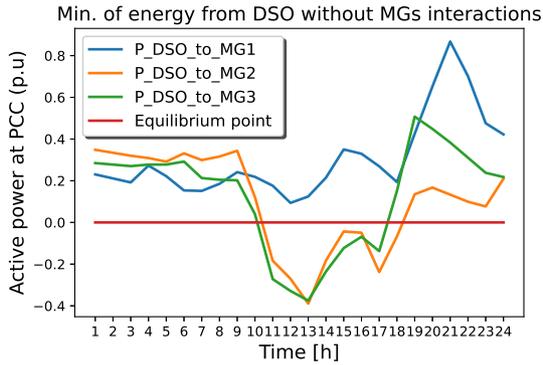}
    	\caption{PDS. Pre-dispatch step results.  Blue, orange, and green lines are the energy demand (positive values) and energy surplus (negative values) of the MG1, MG2, and MG3, respectively.}
    	\label{fig:PDS}
    \end{figure}

     The results of the pre-dispatch are shown in Figure \ref{fig:PDS}.  The negative values represent the energy surplus equivalent to the sale energy capacity per MG.  
    
   {The optimal solution of the PDS establishes the new demands and the energy sales capacities (surplus) of the MGs at time $t \in \mathcal{T}$.  The surplus from MG2 and MG3 is between 10 a.m. and 5 p.m., with the maximum surplus power close to 0.4~pu at 1~p.m.  According to the executed EMS, MG2 and MG3 prefer to sell energy instead of self-supplying energy through their ESS in other time zones due to the surplus price.}
   
   {Following PDS, each MG determines its own prices based on DSO energy prices and the demand curve. DSO prices are higher or equal to MGs' prices to stimulate the local market, which encourages transactions among MGs in the energy market. The price for a residential MG (MG1) is directly related to the load shape behavior. In the case of industrial (MG2) and commercial (MG3) MGs, the prices are established based on a combination of load shape behavior and the DSO prices. This simulation's surplus energy cost is three times less than the DSO energy price. This behavior is presented in Figure \ref{fig:prices}, which shows the sale prices of the agents. The prices of MG2 have three changes concerning MG3's prices during the day. In the morning hours, the MG3 price is smaller than the MG2 price. From 12 m until 4 p.m., they change their price twice, and finally, after 4 p.m., the MG2 has the lowest price.}
  
  {The roles of the MGs are determined by the existence or not of the surplus. If an MG has excess energy, it is considered a seller. If not, it is considered a buyer. The role is assigned by the market framework intertemporally, but it depends on the result of the PDS. The simulation results of PDS show MG1 has the buyer role, while MG2 and MG3 are the sellers. Thus, the next step is the analysis of the interactions among agents according to their roles and capabilities in the ETS step.}

 \subsection{ETS simulation cases} \label{S5-Interact}
  In the energy transactions step (ETS), we introduce  two case sets to analyze the energy interactions based on the relation between the DSO energy price $\zeta_0$ and the surplus price $\Hat{\zeta}_0$.  
    \begin{itemize}
        \item \textbf{Case set C1} $(\zeta_0 = \Hat{\zeta}_0)$: DSO energy price (from DSO to MGs) is equal to the surplus energy price (from MGs to DSO).
        \item \textbf{Case set C2} $(\zeta_0 > \Hat{\zeta}_0)$: DSO energy price to MGs is greater than the surplus energy price (from MGs to DSO).
     \end{itemize}
 The case sets have four sub-cases based on the role of the MG defined in the pre-dispatch (PDS) and the sign of the energy price of the MGs $(\zeta_{e}(t))$ in \eqref{eq:ets}. A positive value of  $(\zeta_{e}(t)$ represents the cost of the buyer, while a negative value represents the revenue of the seller.  
    \begin{itemize}
       \item {\textit{Sub-case 1+}: MG$_2$ sells energy to MG$_1$ and strategy S1 which $\zeta_{e}(t)$ is positive.}
       \item {\textit{Sub-case 2+}: MG$_2$ and MG$_3$ sell energy to MG$_1$ and strategy S1 which means $\zeta_{e}(t)$ is positive.}
       \item {\textit{Sub-case 1-}: MG$_2$ sells energy to MG$_1$ and strategy S2 which means $\zeta_{e}(t)$ is negative.}
       \item {\textit{Sub-case 2-}: MG$_2$ and MG$_3$ sell energy to MG$_1$ and strategy S2 which means $\zeta_{e}(t)$ is negative.}
    \end{itemize}

  {Eight cases are simulated, two for set C1 and four for C2. The cases' names are defined by combining the case set and the sub-case names. Then, for example, in C1.1-, the case C1 is in combination with the sub-case 1- in which $\zeta_0 = \Hat{\zeta}_0$, MG2 sells energy to MG1, and the revenue of MG2 is considered. The sub-cases have two energy providers: DSO and MG2 from 10 a.m. to 5 p.m., while sub-cases 2 have three: DSO, MG2, and MG3 in the same interval of time as sub-cases 1. MG1 is only the buyer in both sub-cases at all time intervals.}

      \subsection{Interactions results}
    
    {ETS is a multi-microgrid energy management system that is described by the model in \eqref{eq:ets}. The new demand and the energy surplus of each MG per hour and roles are the ETS's inputs. The framework also considers the inputs taken into account in the PDS and the energy costs of DSO and MGs.}  The values of the objective function for the ETS cases are given by Table \ref{tab:OFV}.  
    \begin{table}[width=.85\linewidth,cols=3,pos=ht]
    \caption{Objective functions values for the energy transactions step (ETS). The OFV for PDS is 29.6984. }\label{tab:OFV}.
        \begin{tabular*}{\tblwidth}{@{} LL@{} }
            \toprule
           Case & OFV \\
            \midrule
               C1.1+, C1.2+   & 29.6984  \\
               C1.1- & 28.0925  \\
               C1.2- & 27.4943  \\
               C2.1+ & 32.6125  \\
               C2.2+ & 32.6073  \\
               C2.1- & 29.9646  \\
               C2.2- & 28.7322  \\
         \bottomrule
        \end{tabular*}
    \end{table}
  
  {In the optimal solution, C1.2-, which favors increasing seller revenue, has the shortest value, or 27.5 cost units.  In contrast, case C2.1+, which employs the strategy of reducing buyer costs, has the objective function's maximum value, i.e., 32.61 cost units.  As more MGs competitors enter the market, there is a noticeable tendency for the aim function to minimize costs. This finding is reasonable because MGs can select the optimal price based on the established offer for each time period.}

   \subsubsection{Case set C1 ($\zeta_0 = \Hat{\zeta}_0$)}
  
\begin{figure}
    	\begin{subfigure}[b]{0.5\textwidth}
         \centering
         \scalebox{0.5}{\input{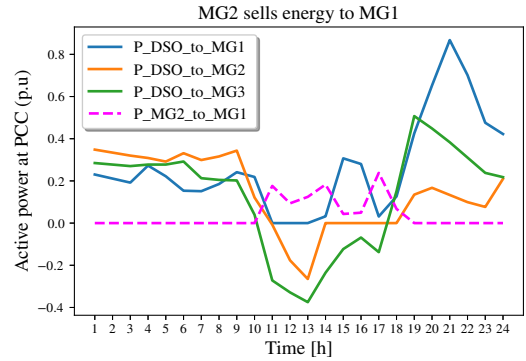}}
         \caption{Case C1.1-}
         \label{fig:C1.1}
     \end{subfigure}
     \begin{subfigure}[b]{0.5\textwidth}
         \centering
         \scalebox{0.5}{\input{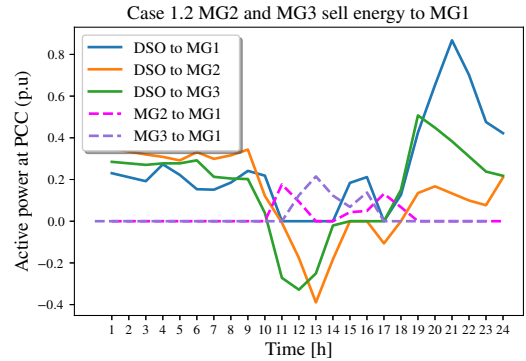}}
         \caption{Case C1.2-}
         \label{fig:C1.2}
     \end{subfigure}
    	\caption{Energy interactions in the cases C1.  Blue, orange, and green lines represent the energy demand (positive values) and energy surplus (negative values) of the MG1, MG2, and MG3, respectively.  The dashed line in magenta depicts the energy sold from MG2 to MG1, while the dashed line in purple depicts the energy sold from MG3 to MG1.}
    	\label{fig:C1}
    \end{figure}

  Figure \ref{fig:C1} shows the energy transactions of the case set C1 composed of two sub-cases, C1.1- and C1.2-, in which the market strategy is based on {increasing } the sellers' MGs revenue. Cases C1.1+ and C2.2+ consider the same energy prices from DSO and MGs surplus, and the result is no interactions between the MGs or variations in the objective function.  In this case, the result is the same as Figure \ref{fig:PDS}.

  Case C1.1- in Figure \ref{fig:C1.1}  shows energy transactions between MG1 and MG2 from 10 a.m. to 6 p.m. The MG1 demand between 11 a.m. and 1 p.m. is fully satisfied by MG2 energy. From 2 p.m. to 6 p.m., the MG2 delivers all available sales energy, reducing the demand for MG1. In both instances, MG1's operational cost is reduced because the energy required from MG1 to the DSO is less than that required in pre-dispatch.

  In case C1.2-, a competitor appears on the market, MG3,  as seen in Figure \ref{fig:C1.2}, in which the sold energy from MG2 to MG1 decreases and MG3 sales increase. Thus, MG1's demand was reduced compared with Case 1.1-. The increased energy availability in the market causes the amount of energy purchased from the DSO to decrease.   

  Recall that in Cases C1.1- and C1.2-, the strategy is based on the seller's benefit. Thus, MG1 (the buyer) makes its energy transaction with the MG that offers the highest price. This strategy aims to maximize the joint profit of the agents interacting in the market. Then, analyzing the dynamics of the energy transactions concerning the market price in Case 1.2-, it can be seen that MG2 is the first agent to sell energy, doing so exclusively between 10~a.m. and 11 a.m. This behavior follows because MG2 has the highest price in the market until 11 a.m. Subsequently, MG3 generates exclusive sales until 2 p.m. Between 2 p.m. and 5 p.m., both MGs sell energy to MG1 since it is the point at which MG3 delivers all its sales capacity. Finally, MG2 completes its sale transaction at 6 p.m. when its sale capacity ends.

    \subsubsection{Case set C2 ($\zeta_0 > \Hat{\zeta}_0$)}

  Case set C2 comprises four cases, two based on the cost strategy (C2+) and the other two on the revenue strategy (C2-). However, the energy surplus from MGs to DSO is considered for both strategies. The common feature of the C2 cases is that the surplus energy price is less than the DSO energy price. 

  The C2+ cases have a strategy that favors the buyers, while the C2- cases benefit the sellers. Additionally, the C2+ and C2- cases have two sub-cases: when MG2 sells energy to MG1 (C2.1+ and C2.1-) and MG2 and MG3 sell energy to MG1 (C2.2+ and C2.2-). The energy transactions for C2.1+ and C2.2+ sub-cases are shown in Figures \ref{fig:C2.1+} and \ref{fig:C2.2+}, and for the C2.1- and C2.2- sub-cases in Figures \ref{fig:C2.1-} and \ref{fig:C2.2-} respectively.
      \begin{figure}
    	\begin{subfigure}[b]{0.5\textwidth}
          \centering
          \scalebox{0.5}{\input{figs/pgf/MGs_C21+.pgf}}
          \caption{Case C2.1+}
          \label{fig:C2.1+}
        \end{subfigure}
        \begin{subfigure}[b]{0.5\textwidth}
          \centering
          \scalebox{0.5}{\input{figs/pgf/MGs_C22+.pgf}}
          \caption{Case C2.2+}
          \label{fig:C2.2+}
        \end{subfigure}
        \caption{Energy interactions in the cases C2.  Blue, orange, and green lines represent the energy demand (positive values) and energy surplus (negative values and dashed lines) of the MG1, MG2, and MG3, respectively.  The dashed line in magenta depicts the energy sold from MG2 to MG1, while the dashed line in purple depicts the energy sold from MG3 to MG1.}
    	\label{fig:C2+}
        \end{figure}

   When the interactions in the C2 cases are examined, it is clear that the market strategy that allows for more energy interactions benefits the seller. Furthermore, increased rivalry among agents encourages energy transactions. This fact is demonstrated in Figure \ref{fig:C2.2-}, which relates to case~C2.2, where it can be observed that both MG2 and MG3 have energy transactions. In this case, less surplus energy has been generated than in other cases. Furthermore, case C2.2 has the lowest objective function value of cases C2, as shown in Table \ref{tab:OFV}.
      \begin{figure}
        \begin{subfigure}[b]{0.5\textwidth}
          \centering
          \scalebox{0.5}{\input{figs/pgf/MGs_C21-.pgf}}
          \caption{Case C2.1-}
          \label{fig:C2.1-}
        \end{subfigure}
        \begin{subfigure}[b]{0.5\textwidth}
          \centering
          \scalebox{0.5}{\input{figs/pgf/MGs_C22-.pgf}}
          \caption{Case C2.2-}
          \label{fig:C2.2-}
        \end{subfigure}
    	\caption{Energy interactions in the cases C2-.  Blue, orange, and green lines represent the energy demand (positive values) and energy surplus (negative values and dashed lines) of the MG1, MG2, and MG3, respectively.  The dashed line in magenta depicts the energy sold from MG2 to MG1, while the dashed line in purple depicts the energy sold from MG3 to MG1.}
    	\label{fig:C2-}
    \end{figure}

  An interesting phenomenon occurs in {case } sets C2, where the MGs with the seller role prefer to use the available energy in their consumption as a consequence of the reduction in the price of the energy surplus $(\Hat{\zeta}_0)$. Figures~\ref{fig:C2+} and~\ref{fig:C2-} show good examples of this fact, where both MG2 and MG3 reduce their demand using their generation resources by themselves, reducing the amount of energy required for the DSO during the morning hours. In case C2.2-, only the MG2 gives a surplus at 1 p.m., the time with the maximum selling capacity due to their PV generation.

  \subsection{Additional results}
  Next, we show additional results analysis on the State of Energy (SoE) of the batteries, the nodal voltages, the nodal powers, and the nodal harmonic distortion of the MGs.

    \subsubsection{Batteries state of energy} 
  This subsection presents results from the SoE of the batteries per MG. Each MG behaves according to the market role, energy prices, and generation capacities.

\begin{figure}[ht]
        \centering
        \includegraphics[scale=0.4]{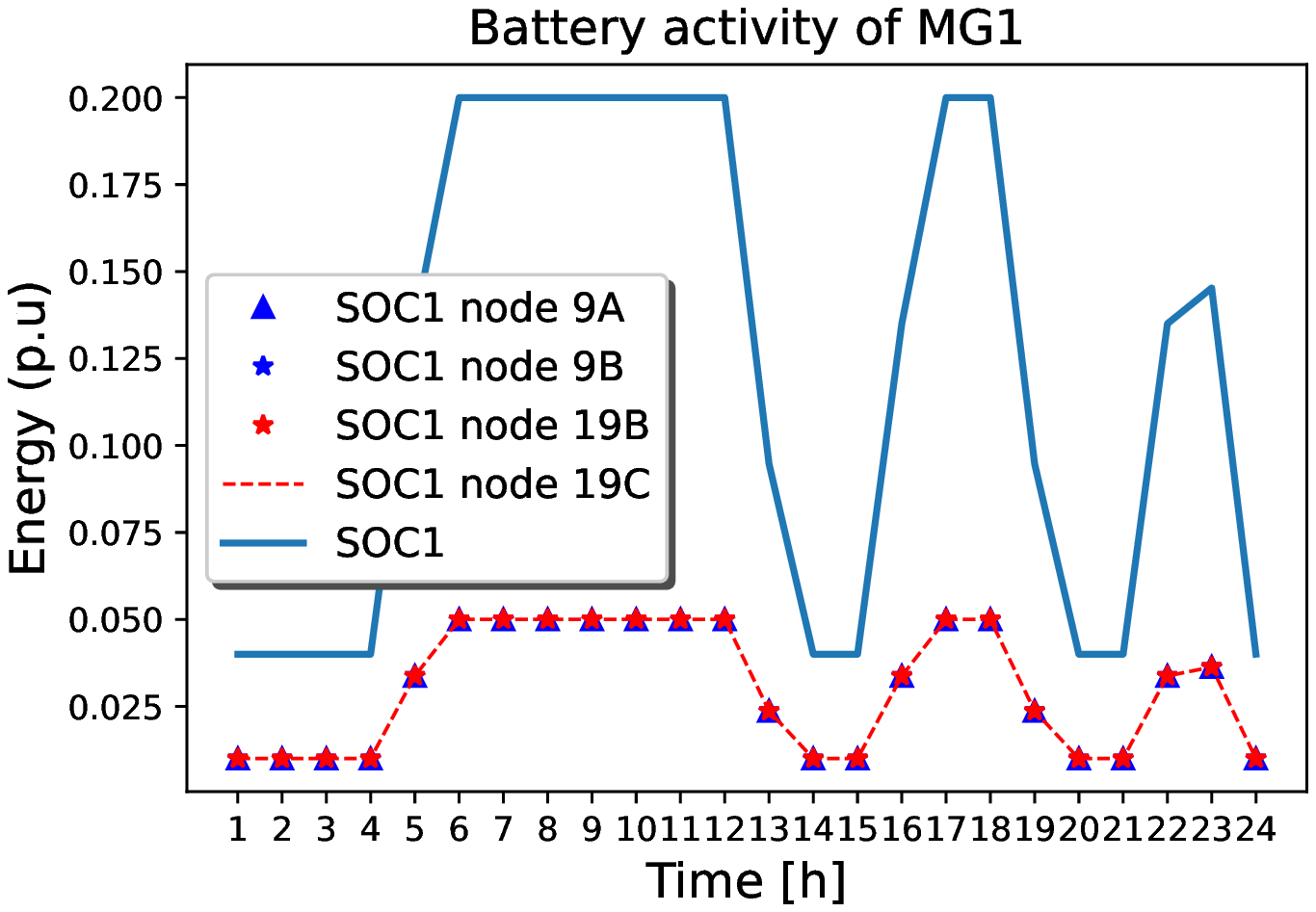}
        \caption{SoE batteries for residential MG. MG1 has two nodes with batteries, node 9 in phases a and b and node 19 in phases b and c. Phase a is identified by triangles, phase b with starts, and phase c with dashed lines. Node 9 is plotted with blue triangles and starts, and node 19 is represented with stars in red and dashed lines. The solid blue line represents the summation of the SoE in the MG1 for all nodes and all phases.}
        \label{fig:SoC_MG1}
    \end{figure}

  In all cases, the MG1 has the same SoE curve as shown in Figure \ref{fig:SoC_MG1}, where it is clear that the battery discharges begin when energy prices are at their highest, i.e., at 12 m, 6 p.m., and 11 p.m. Furthermore, the curve behavior remains consistent in the current cases because MG1 consumes all of its energy resources, as shown in Figure \ref{fig:PDS}.

\begin{figure}[ht]
        \centering
        \begin{subfigure}[b]{0.22\textwidth}
          \centering
           \includegraphics[scale=0.27]{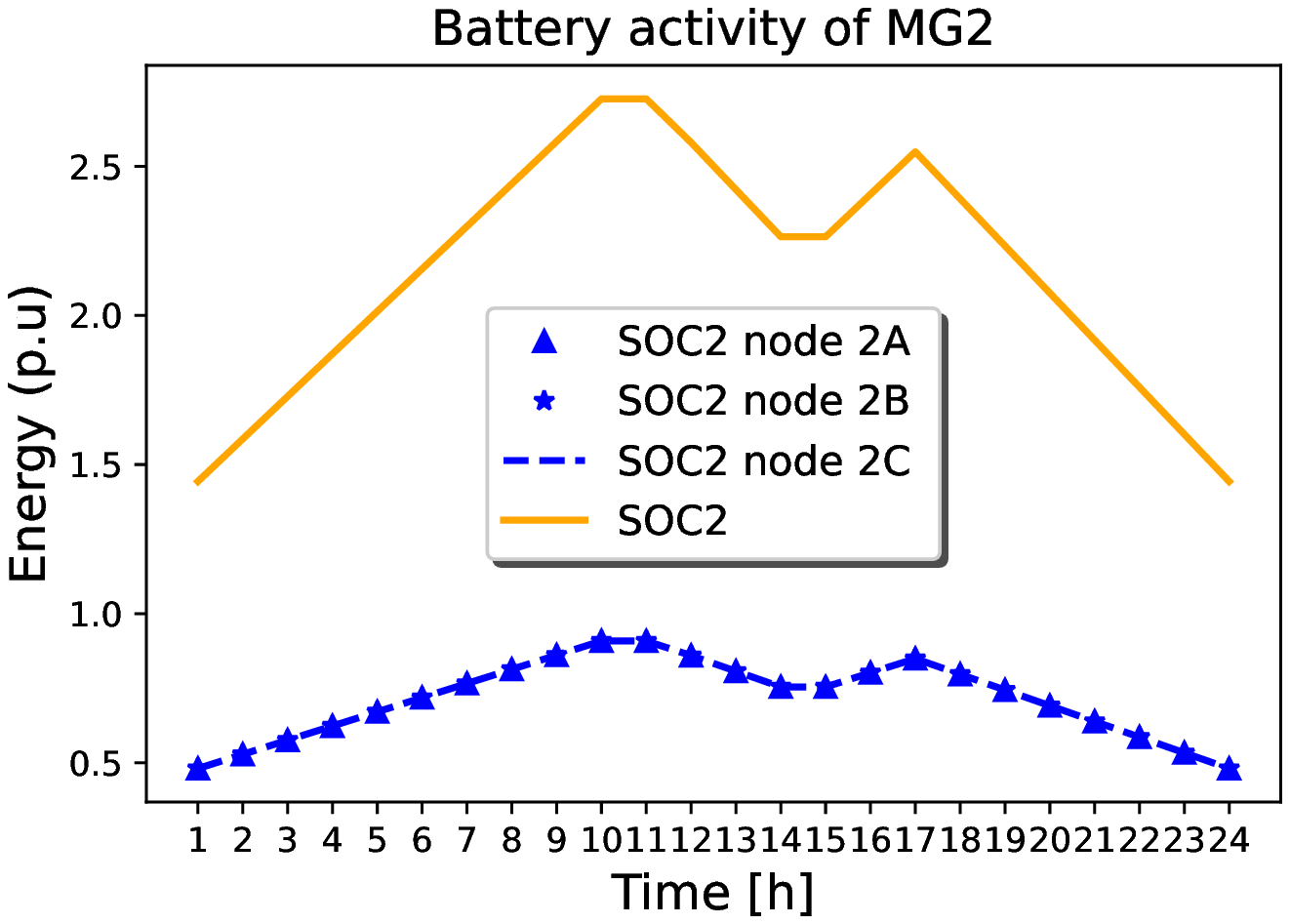}
          \caption{SoE of MG2 Cases C1}
          \label{fig:SoC_MG2_C1.2}
        \end{subfigure}
        \begin{subfigure}[b]{0.25\textwidth}
          \centering
          \scalebox{0.27}{\input{figs/pgf/Batt2_C2.1-.pgf}}
          \caption{SoE of MG2 Case C2.1-}
          \label{fig:SoC_MG2_C2.1-}
        \end{subfigure}
        \begin{subfigure}[b]{0.5\textwidth}
          \centering
          \scalebox{0.3}{\input{figs/pgf/Batt2_C2.2+.pgf}}
          \caption{SoE of MG2 Case C2}
          \label{fig:SoC_MG2_C2.2+}
        \end{subfigure}
        \caption{SoE batteries for industrial MG.  Node two of the MG2 has a three-phased balanced battery system.  The orange curve shows the summation of the SoE in three phases.}
        \label{fig:SoC_MG2}
    \end{figure}

\begin{figure}
    	\begin{subfigure}[b]{0.22\textwidth}
          \centering
           \includegraphics[scale=0.27]{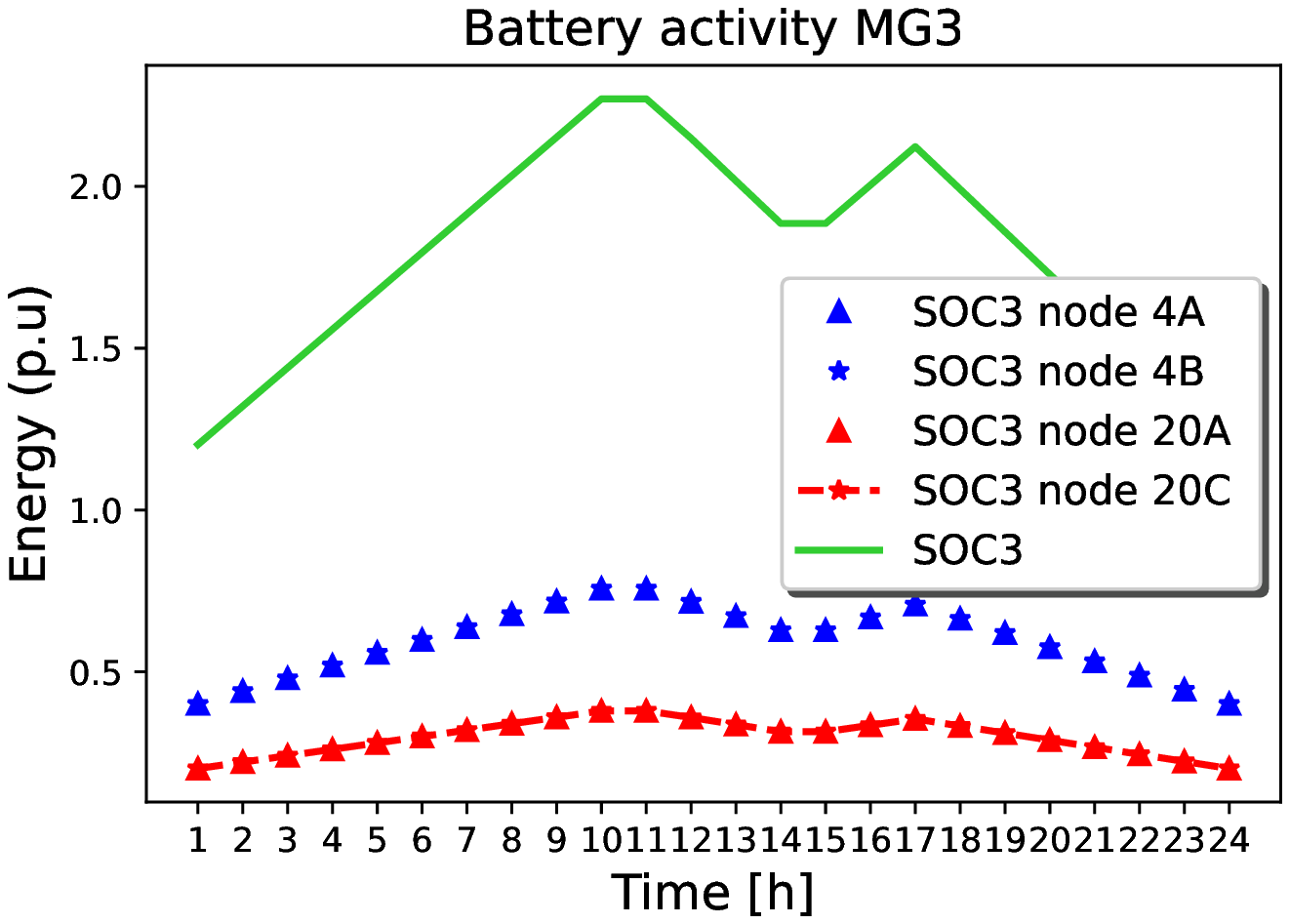}
          \caption{SoE of MG3 \ {Cases } C1.}
          \label{fig:SoC_MG3_C1.2}
        \end{subfigure}
    	\begin{subfigure}[b]{0.22\textwidth}
          \centering
           \includegraphics[scale=0.27]{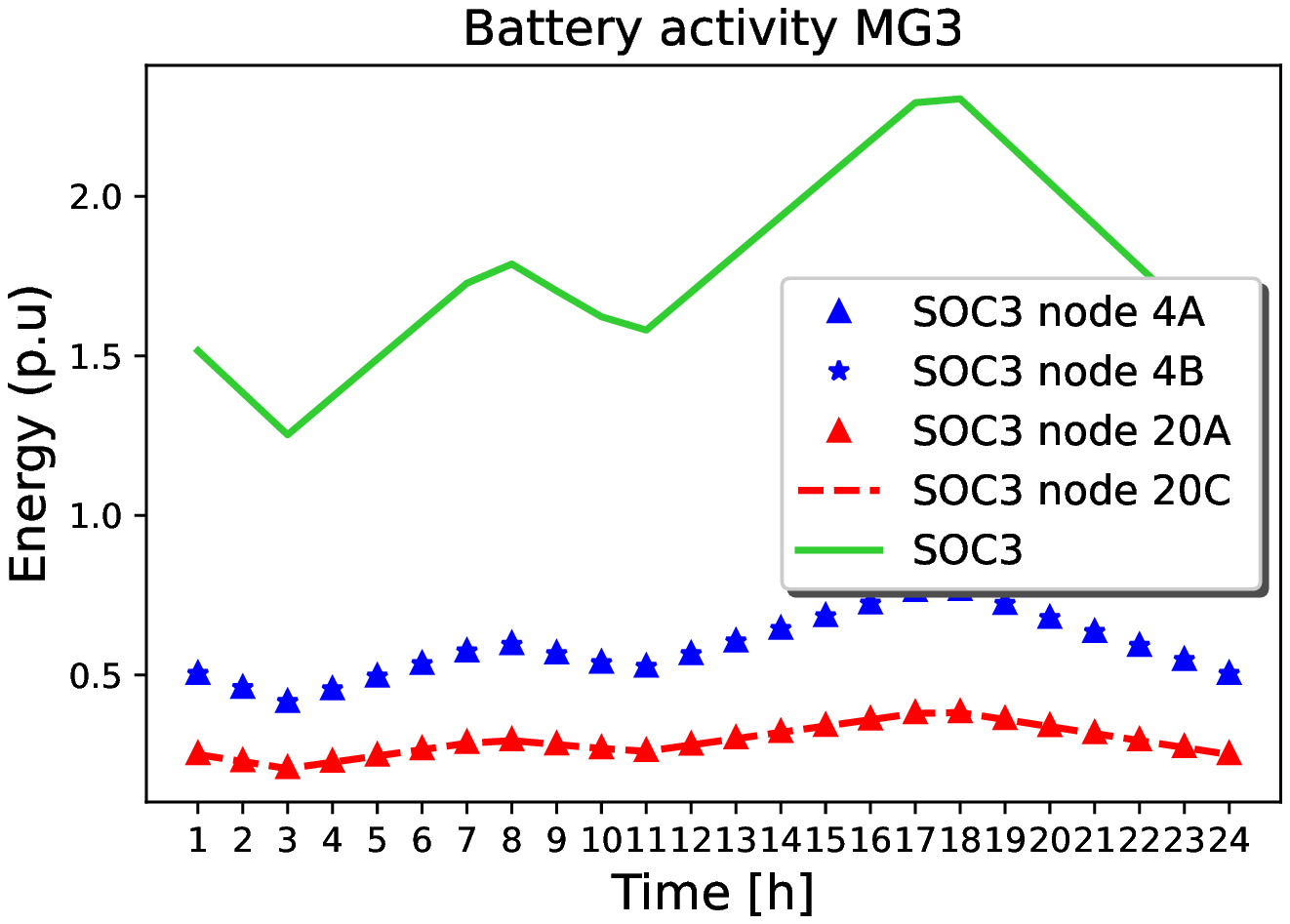}
          \caption{SoE of MG3  {Cases } C2.}
          \label{fig:SoC_MG3_C2.1}
        \end{subfigure}	
    	\caption{SoE batteries for commercial MG. MG1 has two nodes with batteries, node 4 with phases a and b, and node 20 with phases b and c. Phase a is identified with triangles, phase b with starts, and phase c with dashed lines. Node 9 is plotted with blue triangles and starts, and node 19 with red starts and dashed lines. The solid green line represents the summation of the SoE in the MG3 for all nodes and all phases.}
    	\label{fig:SoC_MG3}
    \end{figure}

   MG2 has only one balanced and three-phased Battery Energy Storage System (BESS). For that reason, all phases have the same value of SoE, while MG1 and MG3 have two bi-phased battery systems, one installed at the beginning of the network and the other one at the end of the network. The SoE curves for MG2 and MG3 are shown in Figure \ref{fig:SoC_MG2} and Figure \ref{fig:SoC_MG3}, respectively.

   The SoE for MG2 and MG3 has close behavior because they have similar demand and offer curves, as shown in Figure \ref{fig:PDS}. Furthermore, two representative curves are found, one in Cases C1 and the other in Case C2.   In the Cases C1, for MG2 in Figure \ref{fig:SoC_MG2_C1.2}, and MG3 in Figure~\ref{fig:SoC_MG3_C1.2}, the system charges the batteries during the morning hours to be ready for the energy transactions in which they will receive revenue. After 4 p.m., the BESS charges the batteries again to have the highest energy available at the price peak set by the DSO, where the batteries start to discharge. However, in case  {set } C2, the dynamic changes because the energy required for transactions are less than the demand obtained in the PDS, while in  {case } set C1, the surplus prices are less than the DSO energy price. As a result, the priority for MG2 and MG3 is to use their energy resources at the peak energy price. 

    \subsubsection{Nodal voltages and powers}
\begin{figure}[ht]
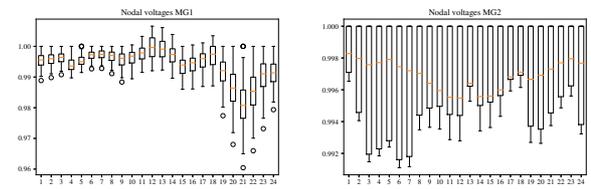
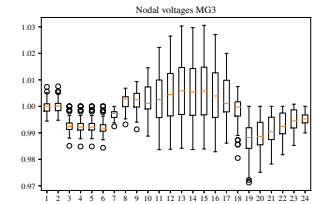

        \centering
        \begin{subfigure}[b]{0.22\textwidth}
          \centering
           \scalebox{0.27}{\input{figs/pgf/V1_box_C2.1+.pgf}}
          \caption{MG1 nodal voltages.}
          \label{fig:V_MG1}
        \end{subfigure}
        \begin{subfigure}[b]{0.25\textwidth}
          \centering
          	\scalebox{0.27}{\input{figs/pgf/V2_box_C2.1+.pgf}}
          \caption{MG2 nodal voltages}
          \label{fig:V_MG2}
        \end{subfigure}
        \begin{subfigure}[b]{0.5\textwidth}
          \centering
          \scalebox{0.30}{\input{figs/pgf/V3_box_C2.1+.pgf}}
          \caption{MG3 nodal voltages}
          \label{fig:V_MG3}
        \end{subfigure}
        \caption{Nodal voltages for residential, industrial, and commercial MGs. Box plots represent the distribution of the nodal voltages per time unit. The boxes include the 25th and 75th percentiles, the orange line corresponds to the distribution median, and the outliers are shown as circles. }
        \label{fig:Volt}
    \end{figure}

  The distribution of the nodal voltage for all phases at time $t \in \mathcal{T}$ is shown in Figure \ref{fig:Volt}. Usually, the nodal voltages in the distribution networks never exceed the value of 1~pu. On the contrary, MGs' voltage increases in some nodes according to the DER location in the network topology. However, according to the standard, the voltage does not exceed the maximum limits allowed $(\delta = 5\%)$ for all MGs.  

   For the MG1, the median of the nodal voltages is below one pu, as shown in Figure \ref{fig:V_MG1}. However, around 50\% of the values of the voltages exceed the 1 pu from 12 m to 3 p.m. This fact is due to the peak in PV generation.  The voltage drops at 9 p.m. due to the demand peak of the MG1.

   In the case of MG2, the nodal voltages have short variations and are in the range of 1 and 0.99 pu.  

   The MG3 has the highest nodal voltage values up to 1 pu, mainly due to the photovoltaic generation and the activity generated by the BESS in the energy transactions between 10 a.m. and 5 p.m. The maximum value reached happens between 1 p.m. and 3 p.m., around 1.03 pu. Additionally, there is a rise in the voltage corresponding to an injection of reactive power by the BESS located in node four from 1 a.m. and 2 a.m., which generates an average power of 0.6 pu for phases $a$ and $b$.

   \subsubsection{Harmonic distortion}

   For all nodes of the MGs, the THD is calculated, and its maximum values are identified, especially at the PCC node.

\begin{figure}[ht]
        \centering
        \begin{subfigure}[b]{0.22\textwidth}
          \centering
           \includegraphics[scale=0.28]{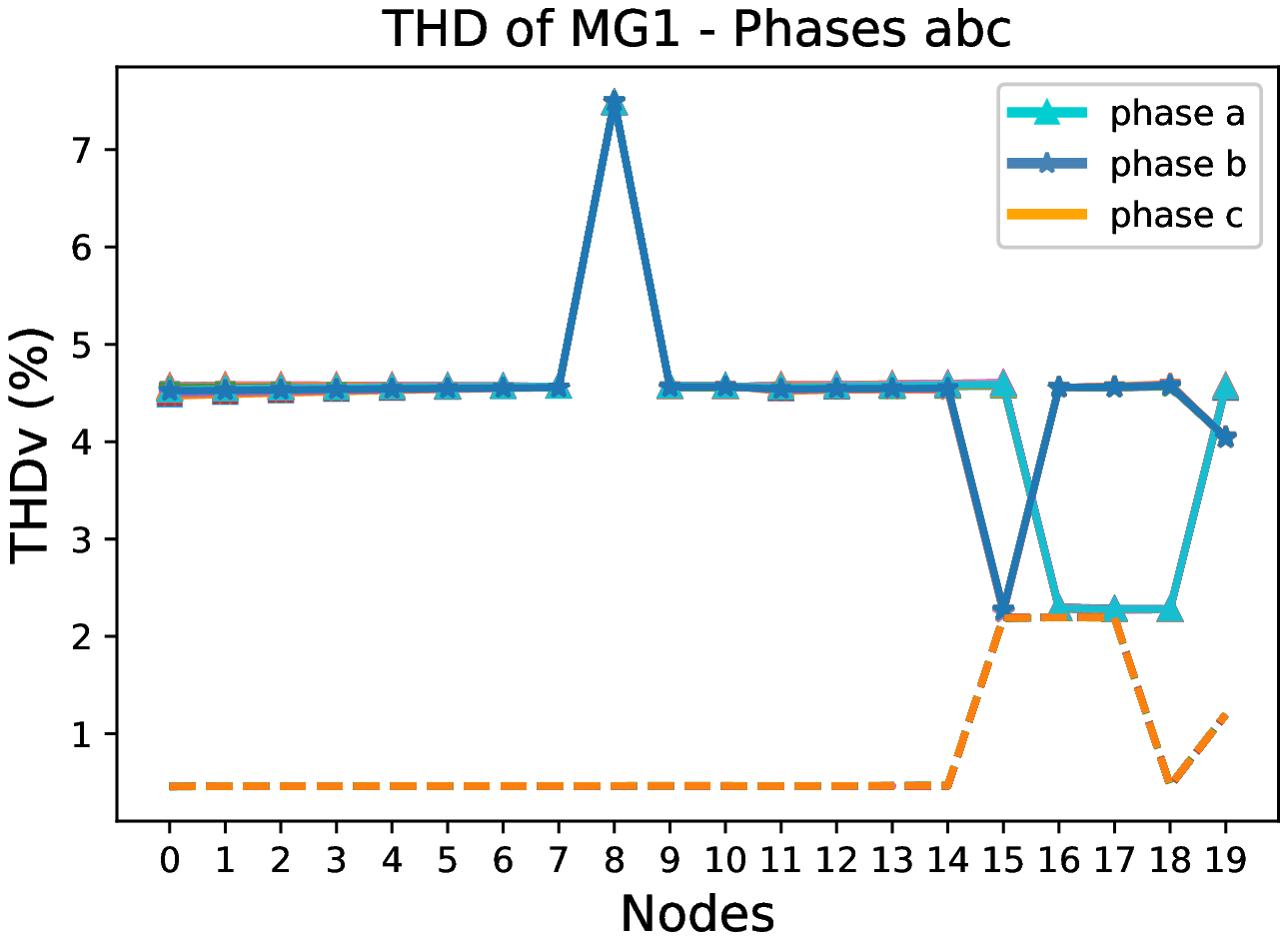}
          \caption{\footnotesize THD MG1}
          \label{fig:THD_MG1_nodes}
        \end{subfigure}
        \begin{subfigure}[b]{0.25\textwidth}
          \centering
          	\includegraphics[scale=0.27]{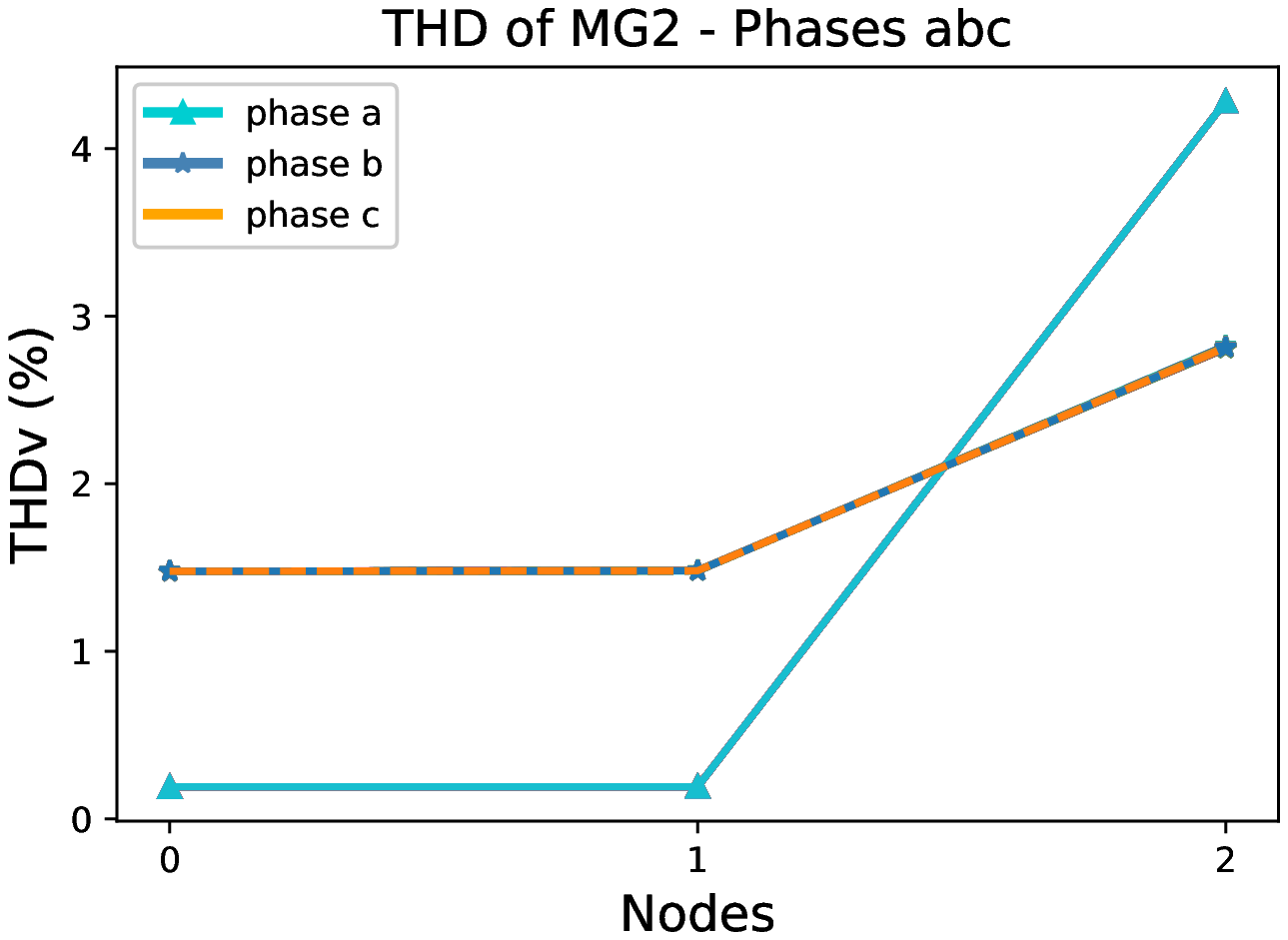}
          \caption{\footnotesize THD MG2}
          \label{fig:THD_MG2_nodes}
        \end{subfigure}
        \begin{subfigure}[b]{0.5\textwidth}
          \centering
          \includegraphics[scale=0.28]{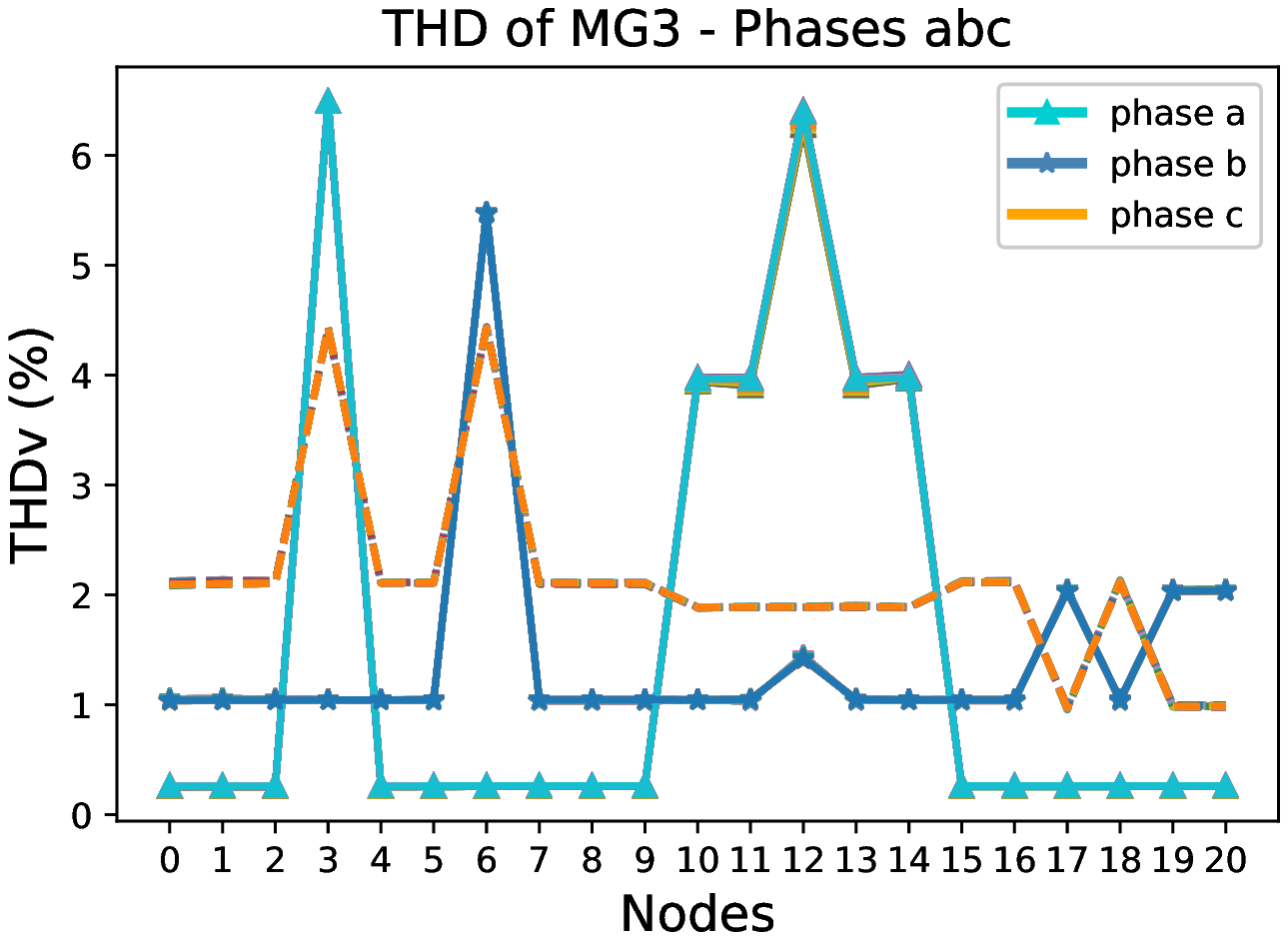}
          \caption{\footnotesize THD MG3}
          \label{fig:THD_MG3_nodes}
        \end{subfigure}
        \caption{THD for MGs per node. Light blue represents phase a, dark blue represents phase b, and dark orange represents phase c.}
        \label{fig:THD_nodes}
    \end{figure}

   Total Harmonic Distortion values per all nodes and time of the MGs are shown in Figure \ref{fig:THD_nodes}. Notably, MG1 has the maximum value of the THD at node 8 in phases $a$ and $b$, which is around 7.5\% for each one. In the same way, MG3 has the maximum value of THD at nodes 3 and 12 at phase $a$, which is approximately 6.5\%. Similarly, MG2 has a maximum of around 4.4\% at node 2 in phase $a$. In all cases, the THD does not exceed the maximum allowed value according to the IEEE Standard 519-2014 \cite{IEEE519-2014}.

  The flexibility of the THD restriction was evaluated by varying the maximum limit established per node, i.e., $\overline{\text{THD}}_i$, resulting in an upper bound depending on the maximum value of THD for the MGs, which for the analyzed case is around 7.5\%. A smaller value of $\overline{\text{THD}}_i$ generates infeasibility in solving the optimization problem. In this sense, the maximum THD value in the nodes gives the flexibility of the parameter $\overline{\text{THD}}_i$.  Thus, lower values of THD require the use of harmonic compensation techniques.

\begin{figure}[ht]
       \centering
        \begin{subfigure}[b]{0.22\textwidth}
          \centering
           \includegraphics[scale=0.27]{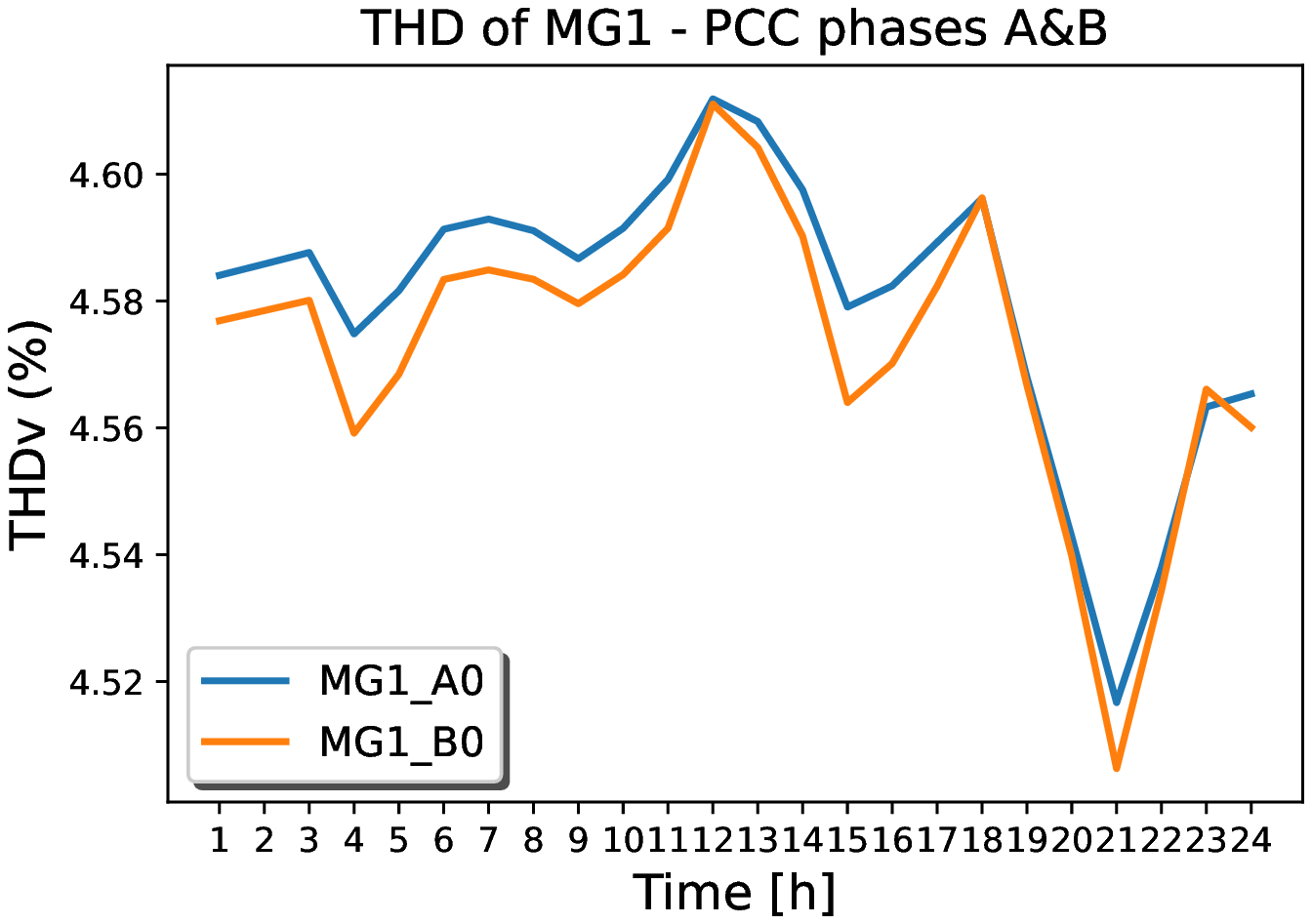}
          \caption{\footnotesize Curve of the maximum values}
          \label{fig:THD_PCC_max}
        \end{subfigure}
         \begin{subfigure}[b]{0.22\textwidth}
           \centering 
            \includegraphics[scale=0.27]{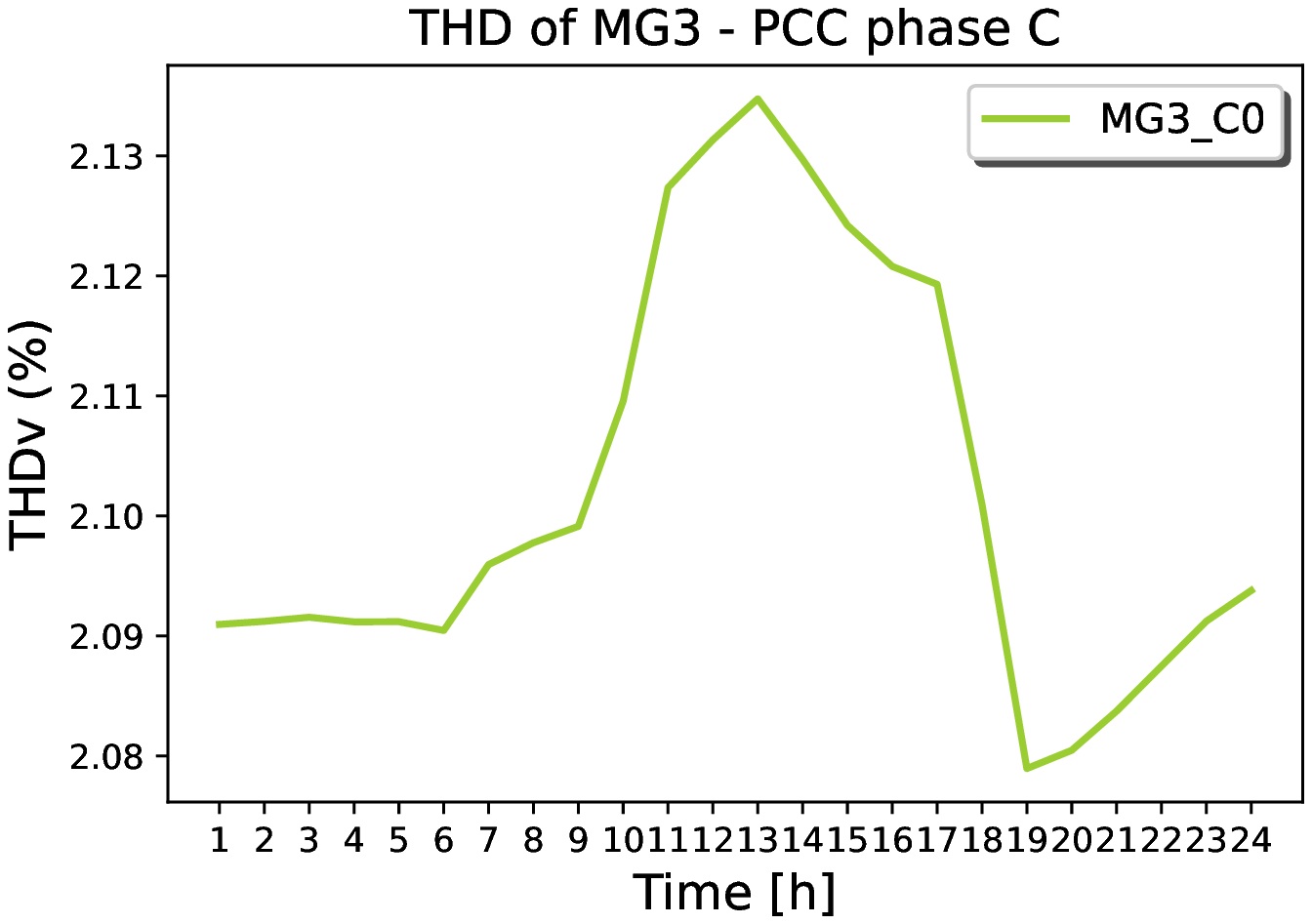}
           \caption{\footnotesize  {Second  highest THD curve}}
          \label{fig:THD_PCC_2max}
        \end{subfigure}
        \caption{THD at PCC node per time{.  (a) shows the curve of maximum values of THD at PPC.  The maximum reached value of THD is less than 4.62\% and occurs at phases A and B of MG1 at noon. (b) shows the curve of the second higher value of THD, which happens at phase C of MG3.} }
        \label{fig:THD_time}
    \end{figure}

  The THD levels at the PCC of each MG by phase are shown in Figure \ref{fig:THD_time}. 
    MG1 has the highest THD levels, with a maximum value of 4.61\% for the phases $a$ and $b$, whose evolution over time is shown in Figure \ref{fig:THD_PCC_max}.  Moreover, the second-highest value is given by the phase $c$ of the MG3 with a 2.3\% of THD as shown in Figure \ref{fig:THD_PCC_2max}.

\section{Conclusion and future works} \label{S7-Conc}

 This paper developed an {MMG energy management system model in a local energy market} with operational and power quality constraints. The proposed model allows for explicit constraints such as power flow, harmonic power flow, power balance, harmonic distortion, voltage limits, and interactions. Interactions occur in the local market and are divided into two steps: pre-dispatch and energy interactions. As a result of the first step, the maximum capacities of energy interactions, demand, and roles (seller and buyer) per time of each MG are obtained. The second step allows us to determine how MGs interact among themselves and determine the DSO's energy surplus.

 {Increasing the number of MGs with seller roles encourages agents to obtain the lowest sum of energy costs for all agents. These facts can be seen in Table \ref{tab:OFV}, where the cases with a higher number of competitors obtain lower values of the objective function. The best option was case 1.2, with a value of 27.49, from the viewpoint of the objective function's minimum value. In this case, the energy and surplus price of the DSO are equal, there are different agents with the role of seller, and the S2 strategy is used. However, from the point of view of incentivizing interactions between MGs, the best option is that the energy prices of the MG agents are lower than those of the DSO. Additionally, in current energy surplus payment schemes for agents selling energy to the grid, this is the most frequent scenario.}
 
 Results showed the market strategy that allows for more energy interactions benefits the seller. Furthermore, increased rivalry among agents encourages energy transactions. As a result of the drop in the price of the energy surplus, microgrids with the seller function prefer to use the available energy for their use. In such circumstances, MGs minimize their demand using their own in-house generation resources, lowering the energy needed for the DSO. 
 
 {The proposed energy management system model of multi-micrigrids interacting in a local market with power quality constraints such as harmonic power flow and harmonic distortion is the first of its kind to the best of our knowledge. The method guarantees the proposed PQ constraints. The nodal voltages and THD limits were met for all three microgrids during all time periods. The voltage value was within the percentage established by the standard, $\pm$ 5\%, being -4\% in the residential MG and 3\% in the commercial MG. The maximum THD value at the common coupling point is 4.61\% for phases a and b of the residential microgrid.}

 Future work should study: local markets, adding more power quality constraints and models of interactions with theories of distributed optimization or game theory, including more DERs, and uncertainties.  Moreover, local markets, such as the reactive or harmonic markets, can be introduced to the model. Moreover, power quality auxiliary services can be offered from MGs to the DSO related to distortion and voltage regulation.  

\section{Acknowledgements} 
Funding: This work was supported by Pontificia Universidad Javeriana and Fulbright Colombia [2021-2022].

\printcredits

\bibliographystyle{cas-model2-names}

\bibliography{references}

\newpage


\bio{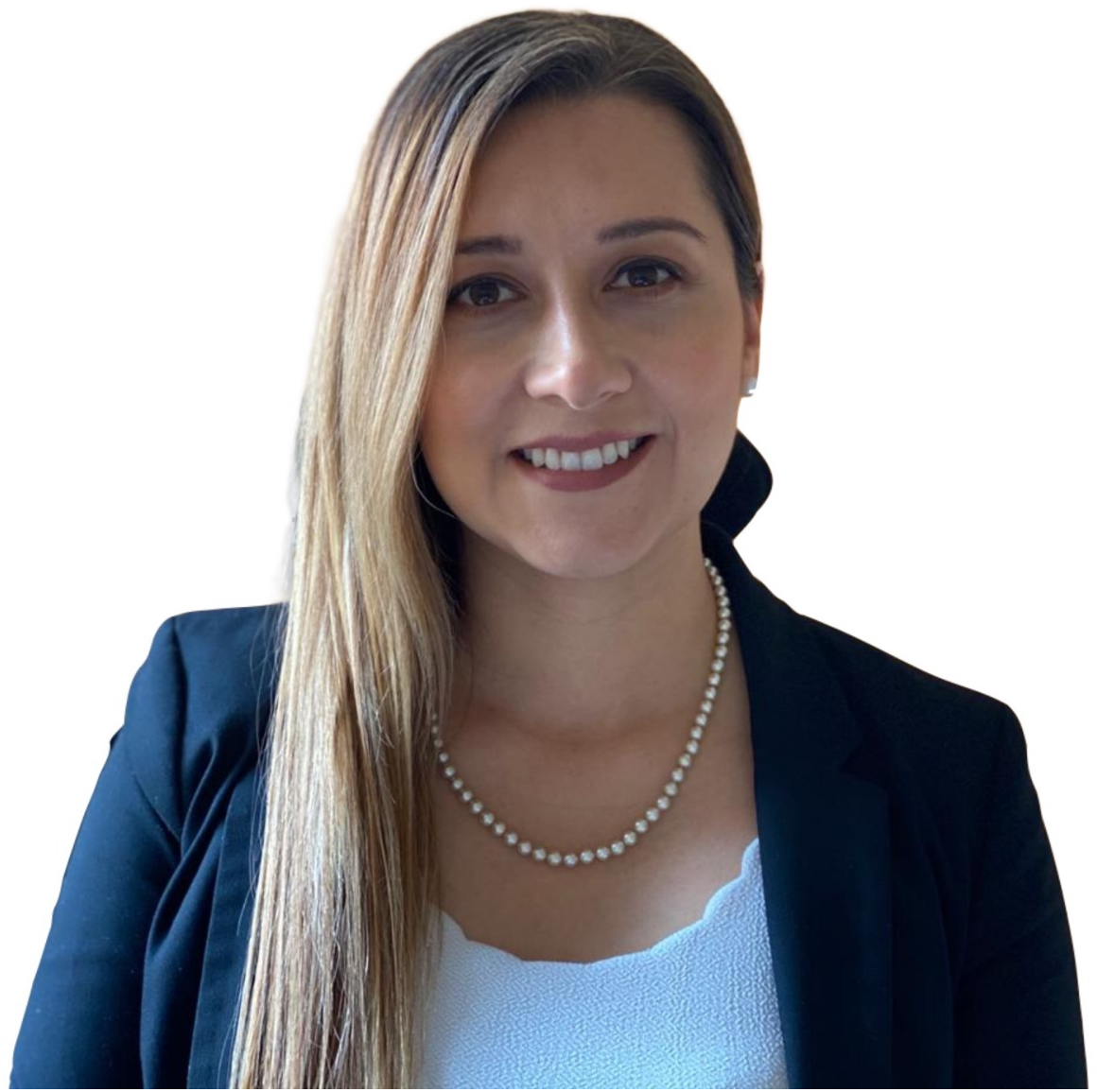}
Johanna Castellanos.
Ph.D. student in Engineering at Pontificia Universidad Javeriana (PUJ) in Colombia and visiting student at Rice University as a Fulbright Scholar in electrical and computer engineering (ECE), 2021-2022.  She received her bachelor's in Mechatronics Engineering from Universidad Militar Nueva Granada and her master's degree in Electronics Engineering at PUJ. She has been a volunteer IEEE member for 15 years.  Her current research interest areas are microgrids, energy management, and optimization.

\endbio

\bio{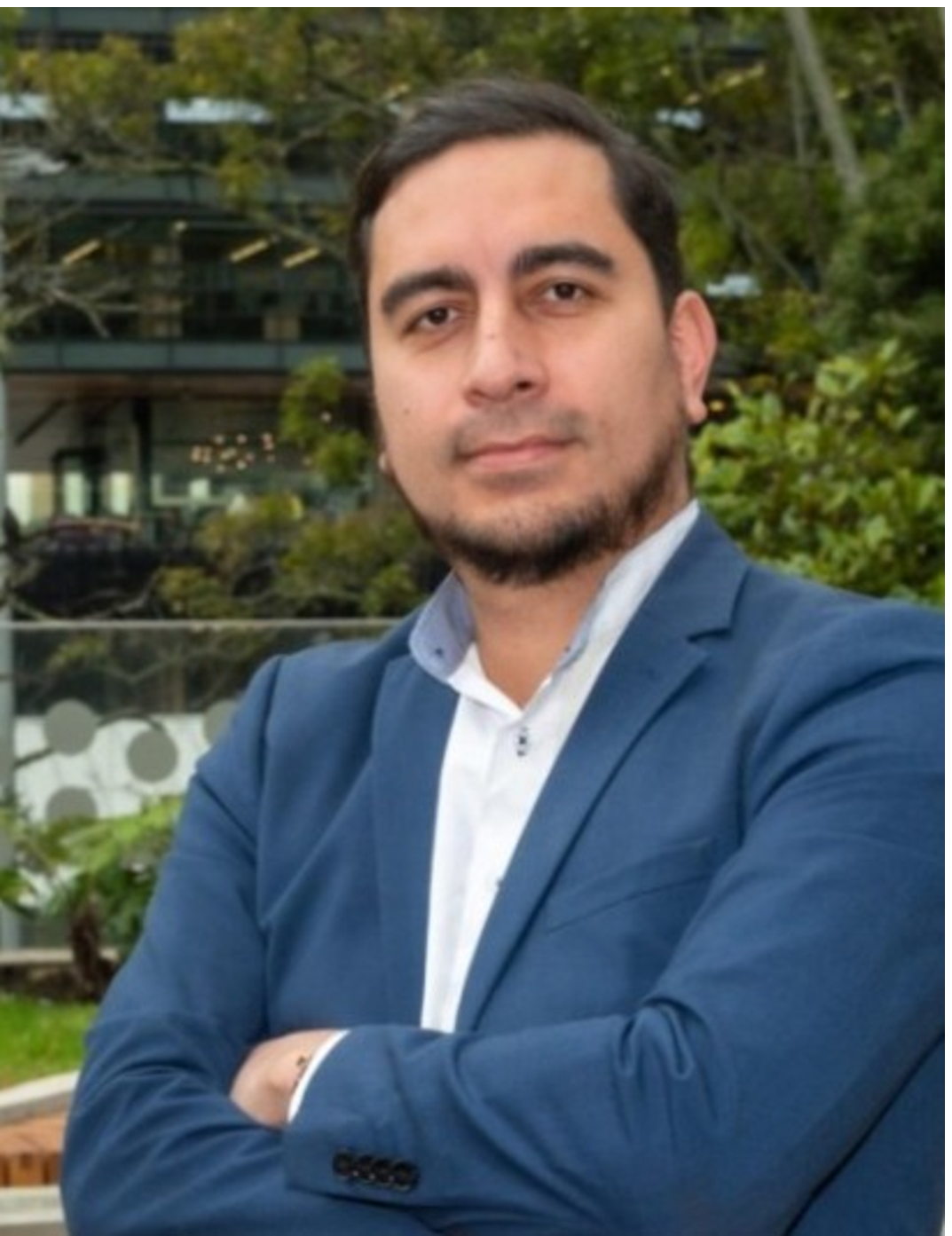}
Carlos Adrian Correa-Florez.
Assistant Professor at Pontificia Universidad Javeriana. He has four years of industrial experience in electrical power systems, over six years as a full-time university professor, and research experience in multiple electrical engineering areas.  He received his bachelor's in Electrical engineering in 2005 and his Master of Electrical Engineering in 2008 from the Universidad Tecnológica de Pereira in Colombia.  In addition, he obtained his Ph.D. in Energy and Processes from MINES ParisTech in 2019.  Areas of interest: renewable energy, optimization models under uncertainty, electrical markets, and smart grids.Ph.D. - MINES ParisTech
\endbio

\bio{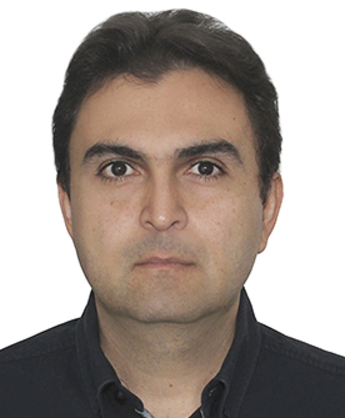}
Alejandro Garcés.
He received his bachelor's and master's degrees from the Universidad Tecnológica de Pereira (Pereira, Colombia, 2006) and his Ph.D. in electrical engineering from the Norwegian University of Science and Technology (Trondheim, Norway, 2012). He is currently an associate professor at the Department of Electric Power Engineering, Universidad Tecnológica de Pereira. He participated in the study Smart Grids Colombia Vision 2030, which defined the roadmap for implementing smart grids in Colombia. Dr. Garces is a senior member of IEEE and a Senior researcher at the National Research System in Colombia. In addition, he is an associate editor in IEEE Transactions on Industrial Electronics and IET-Renewable Power Generation. He also participates in several groups of CIGRE-Colombia and the Colombian chapter of the Society for Industrial and Applied Mathematics (CoSIAM). In 2020 he was awarded the Georg Forster Research Fellowship for Experienced Researchers from the Alexander Von Humboldt Foundation in Germany to continue his research on optimization and control in power systems.
\endbio

\bio{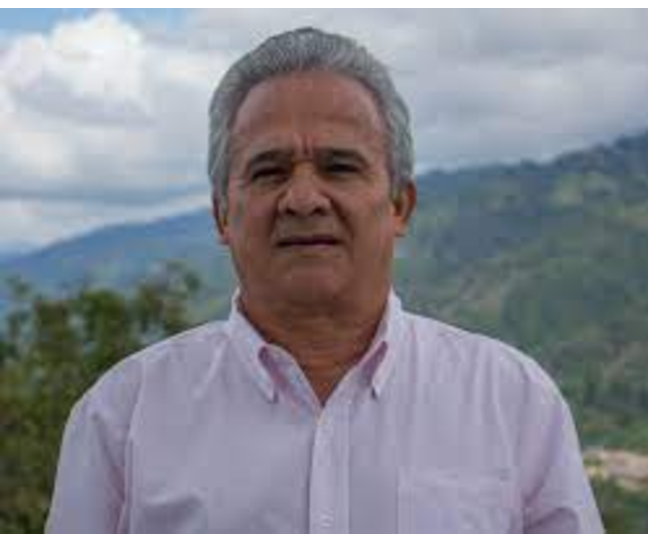}
Gabriel Ordoñez-Plata. He is Titular Professor of the Escuela de Ingenierías Eléctrica Electrónica y Telecomunicaciones at Universidad Industrial de Santander (UIS).  He is a senior researcher and the director of the research group GISEL at UIS.  He received his Doctoral degree in Industrial engineering at Universidad Pontificia Comillas (UPCO), Madrid, Spain.  He obtained his bachelor's degree at UIS. His current work areas are signal processing, electrical measurements, power quality, technological management, and education based on competencies. He is IEEE Senior Member and the IEEE Santanderes Subsection chair.
\endbio

\bio{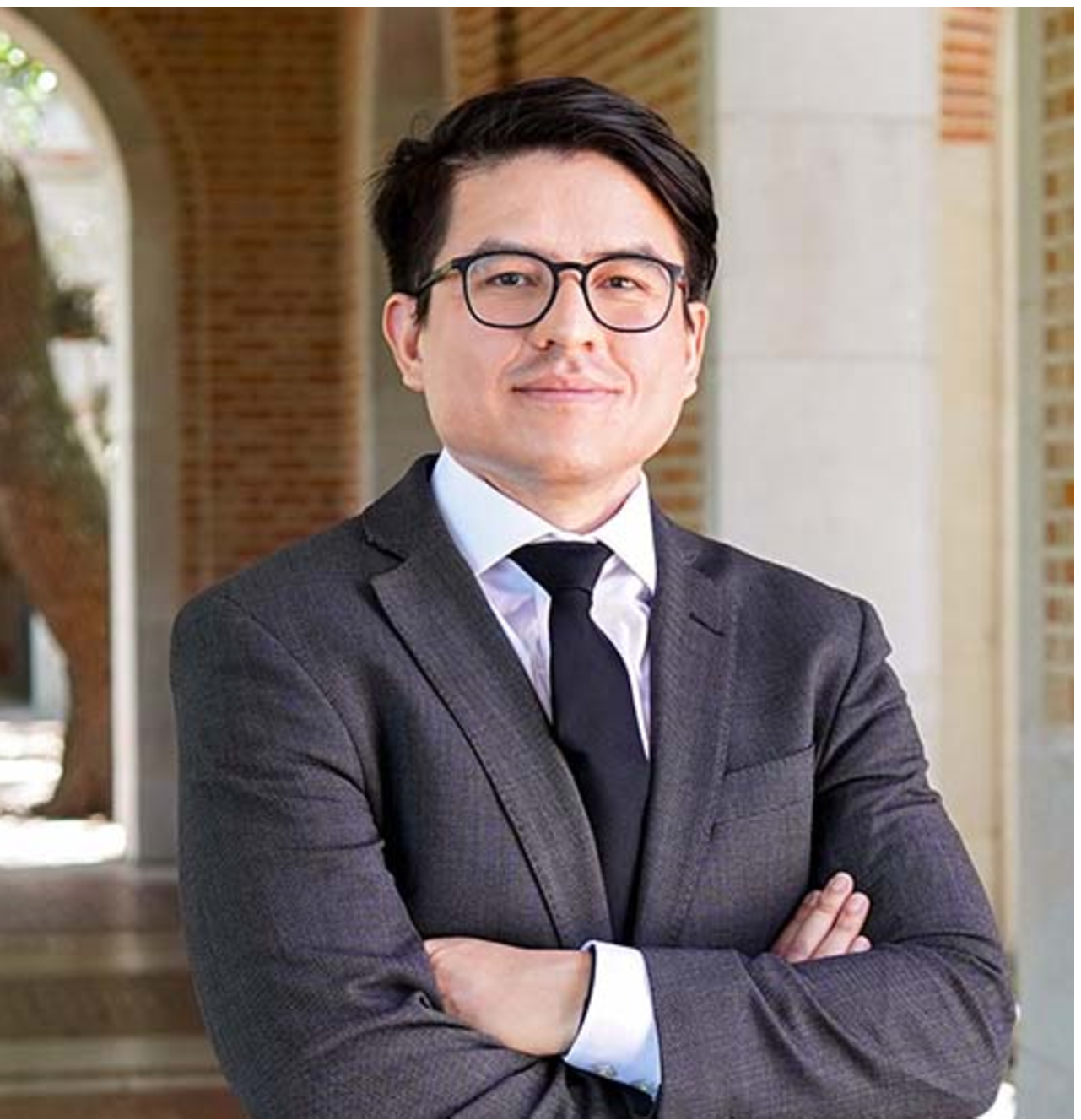}
César A. Uribe received his BSc. in Electronic Engineering from Universidad de Antioquia in 2010. He received an MSc. in Systems and Control from the Delft University of Technology in the Netherlands in 2013. In 2016, be received an MSc. in Applied Mathematics from the University of Illinois at Urbana-Champaign. He continued at the University of Illinois at Urbana-Champaign and, in 2018, received his  Ph.D. in Electrical and Computer Engineering. Urbie was a Postdoctoral Associate in the Laboratory for Information and Decision Systems-LIDS at the Massachusetts Institute of Technology (MIT) and a visiting professor at the Moscow Institute of Physics and Technology. His research interests include distributed learning and optimization, decentralized control, algorithm analysis, and computational optimal transport. He joined Rice ECE Department as Louis Owen Jr., Assistant Professor, in January 2021.
\endbio

\bio{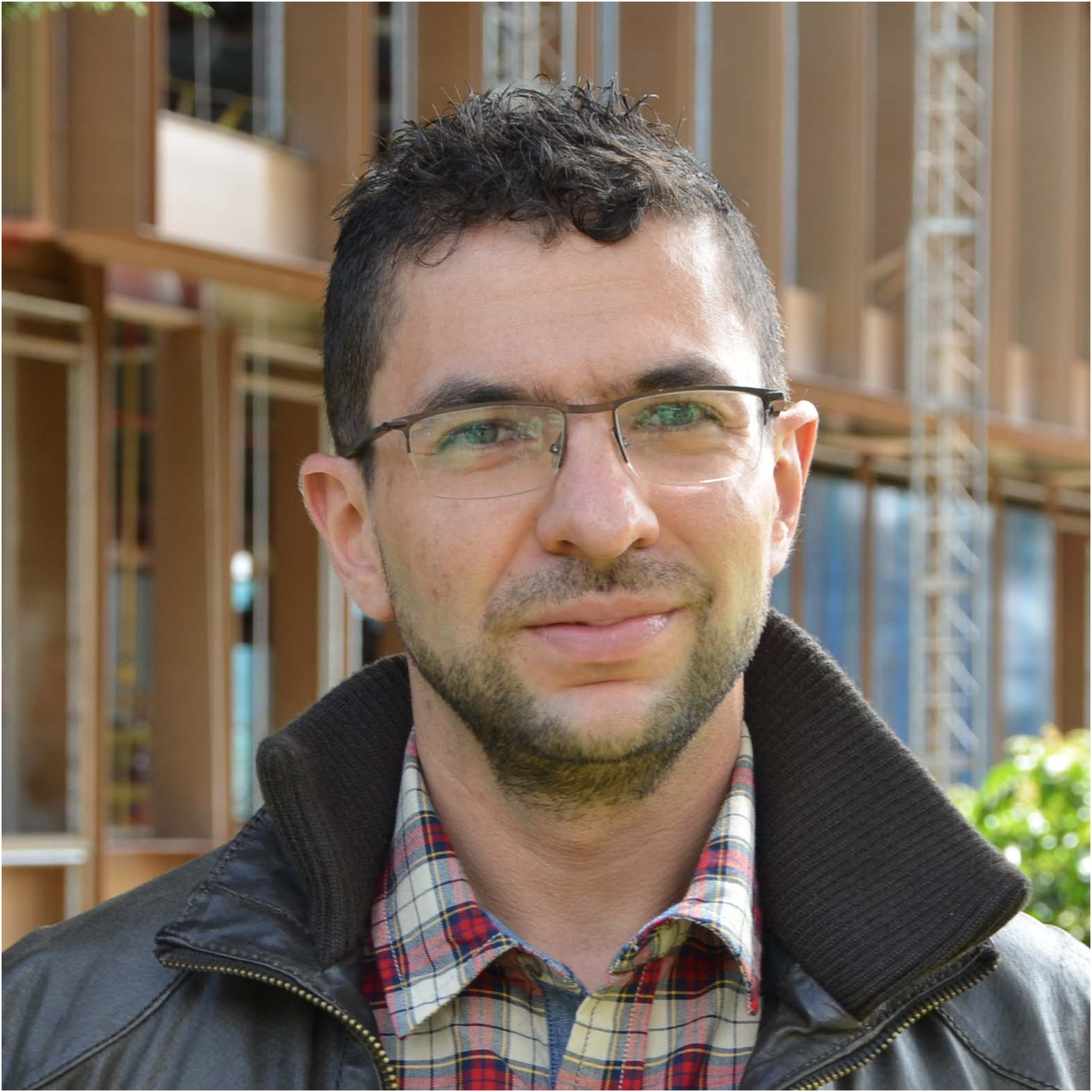}
Diego Patino.
Electronics Department Chief, Engineering Faculty, Javeriana University. He is Electronic Engineer from Colombia Nacional University campus Manizales, Magister in Electronics and Computers Engineering from Los Andes University, and a Doctor in Philosophy in automatic and signal treatment from Institut National Polytechnique de Lorraine. He is a senior researcher, and his research areas are related to control systems, power and energy, smart grids, automation, and others. He has directed research projects on designing and implementing solar energy systems in remote places.
\endbio


\end{document}